\documentclass[a4paper, 12pt]{scrartcl}
  \addtokomafont{section}{\normalsize\selectfont}
  \addtokomafont{subsection}{\normalsize\selectfont}
  \addtokomafont{subsubsection}{\normalsize\selectfont}
  \addtokomafont{paragraph}{\normalsize\selectfont}
  \addtokomafont{caption}{\scriptsize}
  \setkomafont{captionlabel}{\sffamily\bfseries}
  
  \setcapindent{0pt}
\usepackage[hmargin=2cm, top=2.45cm, bottom=2.5cm, footskip=1.25cm]{geometry}
\usepackage[bottom]{footmisc}
\usepackage{booktabs}
\usepackage{tabularx}
  \newcolumntype{Y}{>{\centering\arraybackslash}X}
\usepackage{multirow}
\usepackage[export]{adjustbox}
\usepackage{graphicx}
\usepackage{caption}
\usepackage{subcaption}
\usepackage{amsmath}
\usepackage{amssymb}
\usepackage{enumitem}

\usepackage[backend=biber, style=nature_mod, sorting=none, isbn=false, doi=true, urldate=iso, minbibnames=3, maxbibnames=4, defernumbers=false]{biblatex}
  \DeclareFieldFormat{urldate}{\bibstring{urlseen}\space#1}
  \AtEveryBibitem{\clearfield{month}}
  \AtEveryBibitem{\clearfield{day}}
  \AtBeginBibliography{\small}

\usepackage{xcolor}
\usepackage[colorlinks, urlcolor=blue, citecolor=blue]{hyperref}
\usepackage{subcaption}
\usepackage{url}
\usepackage{setspace}
\usepackage{cancel}
\usepackage{float}
\usepackage{siunitx}
\newcommand{\NumberOfGISAIDCoronaSequences}{14,956,592~}
\newcommand{\NumberOfWholeGenomeSequences}{283,687~}
\newcommand{\LengthOfWholeGenomeSequences}{51,395}
\newcommand{\NumberOfSpikeGeneSequences}{158,190~}
\newcommand{\LengthOfSpikeGeneSequences}{8,237}
\newcommand{\NumberOfOmicronSpikeGeneSequences}{72,309~}
\newcommand{\LengthOfOmicronSpikeGeneSequences}{8,237}
\newcommand{\NumberOfUKSpikeGeneSequences}{25,713~}
\newcommand{\LengthOfUKSpikeGeneSequences}{8,237}

\newcommand{\NumberOfGISAIDInfluenzaHASequences}{24,141~}
\newcommand{\NumberOfGISAIDInfluenzaPBtwoSequences}{23,237~}
\newcommand{\NumberOfHAGeneSequences}{17,223~}
\newcommand{\LengthOfHAGeneSequences}{2,571}
\newcommand{\NumberOfPBtwoGeneSequences}{18,139~}
\newcommand{\LengthOfPBtwoGeneSequences}{2,357}

\newcommand{\NumberOfHIVSequences}{30,964~}
\newcommand{\NumberOfEnvGeneSequences}{30,088~}
\newcommand{\LengthOfEnvGeneSequences}{36,265}

\newcommand{\NumberRunsBenchmarking}{3~}

\extrafloats{100}

\begin{document}

\onehalfspacing

\begin{center}
  \textbf{\large\sffamily Ultrafast topological data analysis reveals pandemic-scale \\ dynamics of convergent evolution}
\end{center}

{\sffamily

\begin{center}

\vspace{2mm}

Michael Bleher\textsuperscript{1*\textdagger}, Lukas Hahn\textsuperscript{1\textdagger}, Maximilian Neumann\textsuperscript{1,2*\textdagger}, Zachary Ardern\textsuperscript{7}, \\
Juan {\'A}ngel Pati{\~n}o-Galindo \textsuperscript{4}, Mathieu Carri{\`e}re\textsuperscript{5}, Ulrich Bauer\textsuperscript{6}, Ra{\'u}l Rabad{\'a}n\textsuperscript{3*}, Andreas Ott\textsuperscript{1*}

\end{center}

\vspace{10mm}

{
\scriptsize
\noindent\textsuperscript{1}Institute for Mathematics, Heidelberg University, Heidelberg, Germany \\
\textsuperscript{2}Institute for Biological Interfaces 5, Microbial Genetics \& Biotechnology, Karlsruhe Institute of Technology, Karlsruhe, Germany \\
\textsuperscript{3}Program for Mathematical Genomics, Department of Systems Biology, Columbia University, New York, NY, USA \\
\textsuperscript{4}Department of Microbiology, Icahn School of Medicine at Mount Sinai, New York, NY, USA \\
\textsuperscript{5}DataShape, Centre Inria d'Universit\'e C\^ote d'Azur, Biot, France \\
\textsuperscript{6}TUM Department of Mathematics and Munich Data Science Institute, Munich, Germany \\
\textsuperscript{7}Wellcome Trust Sanger Institute, Hinxton, United Kingdom
}

\vspace{15mm}

\noindent \textsuperscript{\textdagger}These authors contributed equally to this work.

\vspace{10mm}

\noindent \textsuperscript{*}Corresponding authors: \\
\href{mailto:mbleher@mathi.uni-heidelberg.de}{mbleher@mathi.uni-heidelberg.de} (M.B.) \\
\href{mailto:mn.neumann@kit.edu}{mn.neumann@kit.edu} (M.N.) \\
\href{mailto:rr2579@cumc.columbia.edu}{rr2579@cumc.columbia.edu} (R.R.) \\
\href{mailto:aott@mathi.uni-heidelberg.de}{aott@mathi.uni-heidelberg.de} (A.O.)

\newpage
}

\singlespacing

\section{Abstract}

Genome variants which re-occur independently across evolutionary lineages are key molecular signatures of adaptation.
Inferring the dynamics of such genetic changes from pandemic-scale genomic datasets is now possible, which opens up unprecedented insight into evolutionary processes.
However, existing approaches depend on the construction of accurate phylogenetic trees, which remains challenging at scale.
Here we present EVOtRec, an organism-agnostic, fast and scalable Topological Data Analysis approach that enables the inference of convergently evolving genomic variants over time directly from topological patterns in the dataset, without requiring the construction of a phylogenetic tree.
Using data from both simulations and published experiments, we show that EVOtRec can robustly identify variants under positive selection and performs orders of magnitude faster than state-of-the-art phylogeny-based approaches, with comparable results.
We apply EVOtRec to three large viral genome datasets: \mbox{SARS-CoV-2}, influenza virus A subtype H5N1 and HIV-1.
We identify key convergent genome variants and demonstrate how EVOtRec facilitates the real-time tracking of high fitness variants in large datasets with millions of genomes, including effects modulated by varying genomic backgrounds.
We envision our Topological Data Analysis approach as a new framework for efficient comparative genomics.

\section{Introduction}

Convergent evolution in genomes is a biological signal which often corresponds to variants of high fitness \cite{Sackton2019}.
Recurrently arising variants are of critical importance for instance in the case of microbial pathogens, where adaptive strategies can even be shared across diverse species \cite{Gatt2020, Gutierrez2019, Koonin2022}.
Identifying these genome variants and mapping their evolutionary dynamics is central to understanding pathogen biology and can inform public health responses.
New access to pandemic-scale datasets of millions of genomes creates new opportunities for studying this kind of molecular evolution at an unprecedented scale but also new computational challenges \cite{Hodcroft2021a}.

Experimental approaches consisting of mutating many genome positions \emph{in vivo} and testing the effect on certain phenotypes, are limited by the vast number of possible variations, and also, for instance, the lack of tractable laboratory assays for most viral proteins \cite{Bloom2023a} and technical limitations in implementing unbiased assays at scale \cite{Wei2023}.
Computational methods are well-suited to massive-scale analyses, but face their own challenges \cite{Morel2020}.
These challenges have been addressed by several new approaches, including machine learning approaches \cite{Torada2019, Thadani2023, West2025}, clade-growth approaches \cite{Obermeyer2022, Maher2022, Lee2025}, as well as optimized phylogenetic approaches \cite{Bloom2023a, Zhao2022b, Meijers2023, Haddox2025, Lefrancq2025}.
Phylogenetic approaches have shown good concordance with experimental fitness measurements~\cite{Bloom2023a}.
High fitness variants can be inferred via genomic recurrence \cite{Maeso2012}---if a mutation confers an evolutionary advantage, we expect it to re-occur in independent lineages of the phylogeny and its frequency should increase with time.

However, phylogenetic approaches rely on the construction of accurate phylogenetic trees, which is challenging when the number of genomes exceeds tens of thousands \cite{Hodcroft2021a}.
Common strategies to overcome this difficulty include downsampling \cite{Hadfield2018} and stepwise addition of new sequences onto an existing phylogenetic tree \cite{Turakhia2021, DeMaio2023, Ye2022}. However, downsampling can reduce accuracy through loss of information, and stepwise addition can be used for very large genomic datasets but results in less accurate tree topologies.
An additional conceptual challenge for phylogenetic approaches is that homoplasies and recombination can confuse the construction of phylogenetic trees \cite{Steenwyk2023, Kapli2021, Turakhia2020}.
Such merging of distinct ``vertical'' lineages can arise from several evolutionary scenarios.
If a genome of an organism acquires genetic material from a different genome through horizontal gene transfer or recombination, we will observe that parts of the newly generated genome resemble the ``vertical'' parent, while others resemble the genome of the ``horizontal'' organism that exported  that material \cite{Bjornson2024, Stott2020, Posada2002, Schierup2000}.
Homoplasies are generated  when the same mutation occurs independently twice and is retained in both lineages, making the two sampled strains more similar than expected \cite{Crispell2019}.

Here we present EVOtRec (\url{https://github.com/ottamj/evotrec}), an organism-agnostic, fast and scalable approach based on persistent homology, a method from Topological Data Analysis  \cite{Carlsson2009}, that overcomes limitations of phylogenetic approaches by quantifying the dynamics of convergent evolution solely through its topological footprint in the genomic dataset, without the need to reconstruct a phylogenetic tree.
We analyze the sensitivity and robustness of our method with simulated evolutionary scenarios and apply it to three pandemic-scale viral genome datasets: coronavirus \mbox{SARS-CoV-2}, avian influenza virus A subtype H5N1 and human immunodeficiency virus HIV-1.
We assess EVOtRec’s computational performance in relation to leading phylogenetic tools, and also assess the different tools’ ability to identify fitness-increasing genomic variants detected in previously published experimental data.
Using our new approach, we track potentially fitness-increasing effects of individual genomic variants over time and investigate hotspots of convergent evolution dynamics across the genome, modulated by varying genomic backgrounds.

\section{Results}

\subsection{Topological recurrence analysis with EVOtRec quantifies convergent evolution}

\begin{figure}[h!]
\centering
\includegraphics[width=16cm]{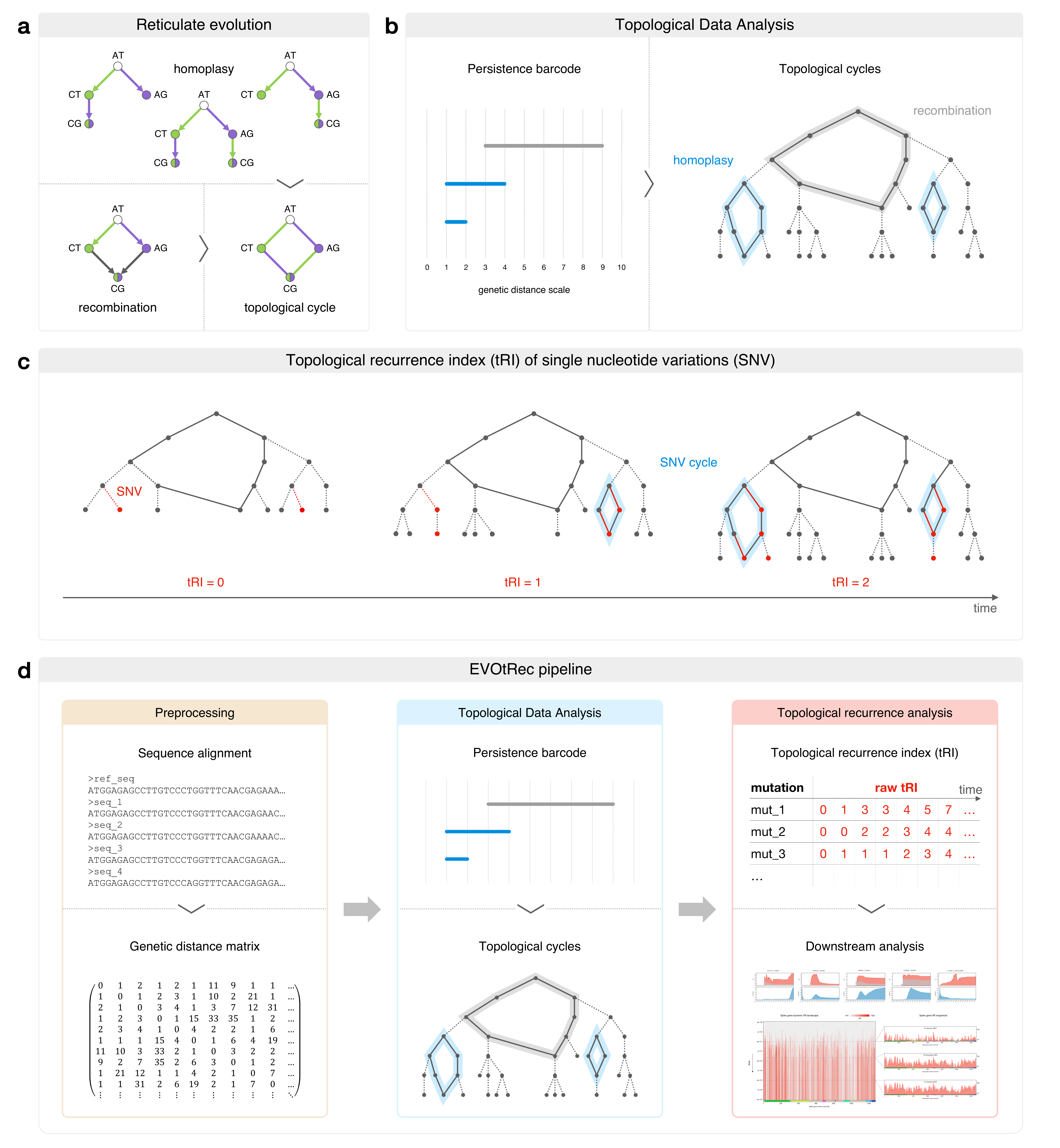}
\caption{\textbf{\sffamily{Topological recurrence analysis with EVOtRec quantifies convergent evolution.}}
\textbf{(a)} Possible evolutionary histories for a dinucleotide sequence.
The coloring of the edges corresponds to a specific mutation, while the coloring of the nodes indicates the presence of the genomic variant resulting from this mutation.
Reticulate events (implying homoplasy or recombination) lead to the presence of four alleles for which there is no single consistent phylogeny. On the genomic level, genetically identical individuals cannot be distinguished and incompatible phylogenies are represented by a topological cycle in the corresponding phylogenetic network.
\mbox{\textbf{(b)}}~Topological Data Analysis detects reticulate events solely by means of their topological footprint in the genomic dataset, without the need to construct a possibly ambiguous phylogenetic tree (\nameref{methods}).
On the technical level, this is achieved by computing a persistence barcode. Each bar in the barcode corresponds to a topological cycle in the reticulate phylogeny, with bars starting at small (large) genetic distance scale typically corresponding to homoplasies (recombination events).
\mbox{\textbf{(c)}}~SNV cycles are topological cycles in which sequences corresponding to adjacent nodes differ by single nucleotide variations (SNV) only, i.e., the genetic distance (edge length) between any two adjacent nodes is $1$.
These cycles correspond to bars starting at genetic distance~$=1$ in the persistence barcode.
The raw topological recurrence index (raw tRI) of a specific SNV is the total number of SNV cycles which contain an edge corresponding to this SNV (\nameref{methods}).
Intuitively, mutations contributing to tRI signals are more likely to happen if the mutation confers some evolutionary advantage, because they appear independently in different lines of descent.
In the displayed phylogeny, the SNV colored in red has a raw tRI of~$2$ as it is acquired in precisely two SNV cycles.
For genome sequence alignments which incorporate temporal information, the time-dependent tRI can resolve topological signals of recurrence at different time steps (\nameref{methods}).
\textbf{(d)} The EVOtRec (topological Recurrence in EVOlution) pipeline (\nameref{methods}) takes as input a sequence alignment (including temporal information) with a fixed reference sequence and returns (time series) tRI data for every SNV in the alignment.
}
\label{fig:DefinitionOftRI}
\end{figure}

\citeauthor{Chan2013} \cite{Chan2013} initiated the use of persistent homology to extract global evolutionary features from genome datasets. This method detects topological cycles in the dataset which correspond to apparent reticulate events (implying homoplasy or recombination) in the phylogeny. Persistent homology captures all these events, and also the scale of the events. All information is compiled into a stable and unbiased descriptor known as a \emph{persistence barcode} (\autoref{fig:DefinitionOftRI} and \hyperref[fig:ripser]{Supplementary \autoref{fig:ripser}}). Bars in the barcode starting at small genetic distance scale typically correspond to homoplasies, while bars starting at larger scale are indicative of recombination events involving the exchange of larger portions of genetic material \cite{Chan2013, Camara2016, Rabadan2019}.

We use persistent homology to introduce a novel index of recurrence that quantifies convergent evolution directly through topological cycles in the genomic dataset, without the need for computationally expensive reconstruction of a phylogenetic tree.
\emph{SNV cycles} in the genomic dataset are topological cycles that consist of a series of genomes that approximates all evolutionary steps as faithfully as possible in terms of single nucleotide variations (SNV).
We define the \emph{topological recurrence index (tRI)} of a specific SNV as the total number of SNV cycles containing this SNV (\autoref{fig:DefinitionOftRI} and \nameref{methods}).
The rationale behind this definition is that mutations observed in an SNV cycle occur independently twice in different lines of descent within this cycle, which implies that the mutation may confer some evolutionary advantage, given that the genomes observed result from processes including filtering by natural selection \cite{Stayton2015, Crispell2019}.
A proportion of tRI signals may also be due to other factors, and the nonuniformity of mutational spectra \cite{Bloom2023b} can bias tRI signals towards particular nucleotide mutation types (\hyperref[fig:tri_fds]{Supplementary \autoref{fig:tri_fds}}).
To compensate for these effects, we implemented several tRI corrections in the topological recurrence analysis (\nameref{methods}).
Most importantly, we normalize the tRI by its mean per nucleotide mutation type across four-fold degenerate sites \cite{Bloom2023a, Bloom2023b, DeMaio2021}.

The topological recurrence analysis can be carried out for sliding windows across the genome (windows of a defined length or single genomic regions), based on sub-alignments of appropriately truncated genome sequences (\autoref{fig:DynamicsOftRI}, \hyperref[fig:tri_region_subalignment]{Supplementary \autoref{fig:tri_region_subalignment}} and \nameref{methods}) \cite{Rabadan2019}.
In this case, mutations outside the window are ignored, which typically leads to the creation of more topological cycles and stronger tRI signals, revealing a more detailed picture of convergent evolution in that region of the genome.

We implemented our method in the EVOtRec (topological Recurrence in EVOlution) pipeline (\url{https://github.com/ottamj/evotrec}).
This pipeline takes as input a sequence alignment with a fixed reference sequence and returns a list of SNVs with corresponding tRI scores (\autoref{fig:DefinitionOftRI}).
EVOtRec is designed for the efficient and scalable analysis of very large genomic datasets, and is based on highly optimized algorithms for the rapid computation of the persistent homology of genomic datasets.
Hammingdist \cite{Keegan2024} is used to compute genetic distances, and Ripser \cite{Bauer2021} is used for the subsequent computation of the persistence barcode and the localization of topological cycles (\autoref{fig:DefinitionOftRI}, \hyperref[fig:ripser]{Supplementary \autoref{fig:ripser}} and \nameref{methods}).
For sequence alignments that integrate temporal information, such as the collection date of genomic isolates, EVOtRec applies the MuRiT algorithm \cite{Bleher2022} to enable a time series analysis of convergent evolution based on time-dependent tRI (\autoref{fig:DefinitionOftRI} and \nameref{methods}).
EVOtRec processes genetic distances in a tailored sparse matrix format which keeps track of short distances only, thereby greatly reducing the complexity of persistent homology computations (\hyperref[fig:ripser]{Supplementary \autoref{fig:ripser}} and \nameref{methods}).
In this way, EVOtRec serves as a useful tool for the real-time monitoring of emerging potentially adaptive genomic variants in ongoing evolutionary processes and for the mapping of convergently evolving genomic variants over time.

\subsection{Simulations show that topological recurrence analysis is sensitive to positive selection and robust to topological noise and stochastic sequencing errors}

\begin{figure}[h!]
  \centering
  \includegraphics[width=\textwidth]{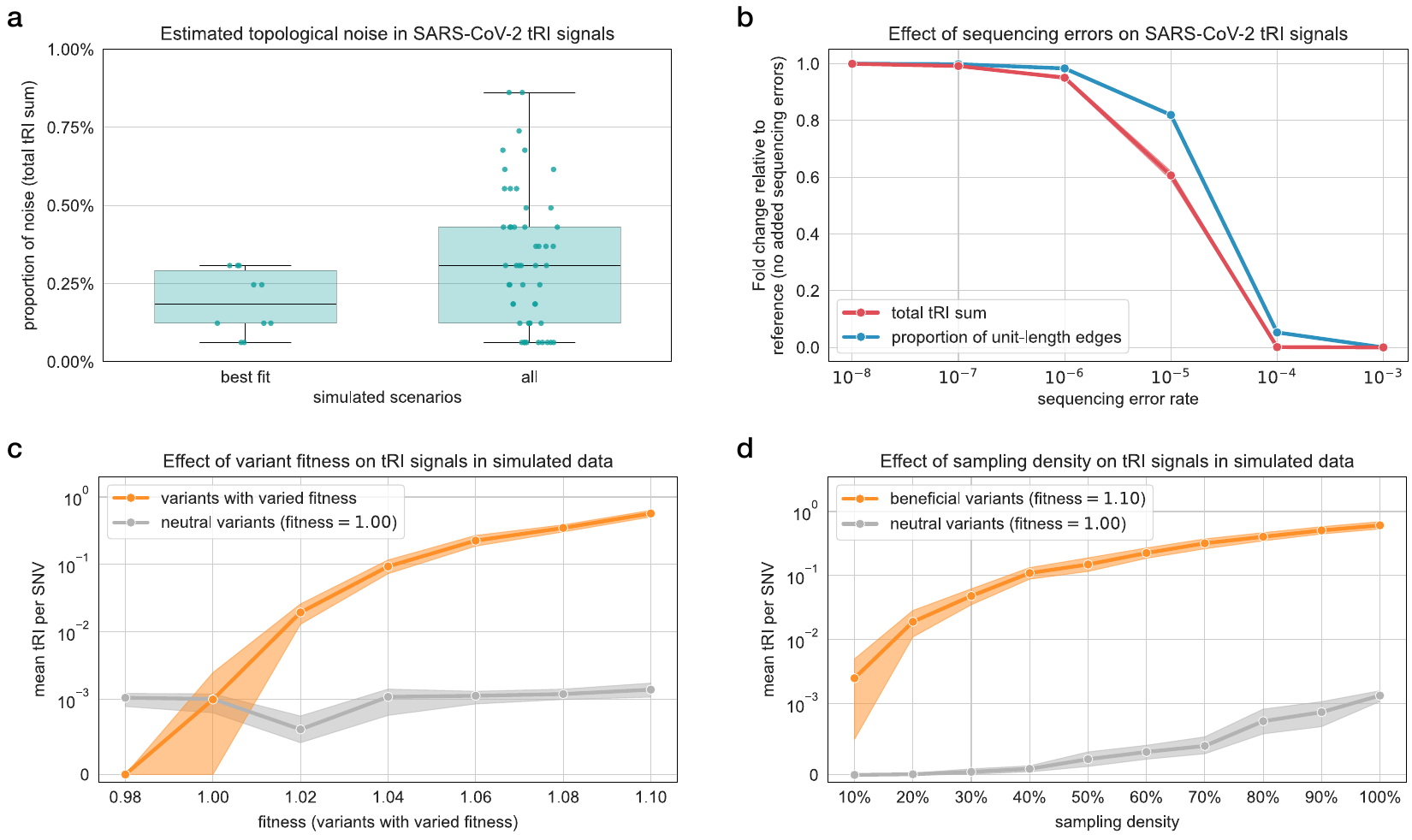}
  \caption{
\textbf{\sffamily{Simulations show that topological recurrence analysis is sensitive to positive selection and robust to topological noise and stochastic sequencing errors.}}
\textbf{(a)} Box plot with overlaid strip plot showing the proportion of background topological noise in tRI (total tRI sum across all variants) of a SARS-CoV-2 whole genome alignment with 100k sequences, estimated from five different simulated neutral evolutionary scenarios (for details see \hyperref[fig:topological_noise]{Supplementary \autoref{fig:topological_noise}} and \nameref{methods}).
\textbf{(b)} We assessed how stochastic sequencing errors affect SARS-CoV-2 tRI signals by adding simulated stochastic sequencing errors to a real SARS-CoV-2 whole genome alignment with 200k sequences, for varying error rates ranging from $10^{-8}$ to $10^{-3}$ (for details see \hyperref[fig:sequencing_errors]{Supplementary \autoref{fig:sequencing_errors}} and \nameref{methods}).
The line plot (mean ± 95\% confidence band) shows the relative change in the total tRI sum across all variants and in the proportion of unit-length edges among all possible edges in the genomic dataset.
Here, relative change refers to the alignment with added sequencing errors compared against the reference alignment without added errors.
We found that the total tRI sum did not increase after adding sequencing errors but instead decreased to zero at higher error rates.
This phenomenon is explained by the fact that higher sequencing error rates increase genomic distances, thus progressively reducing the proportion of unit-length edges which are the building blocks of SNV cycles (\autoref{fig:DefinitionOftRI}).
\textbf{(c-d)}~Dependence of tRI signals on variations in variant fitness and sampling density in simulated synthetic alignments (for details see \hyperref[fig:tri_simulated_scenarios]{Supplementary \autoref{fig:tri_simulated_scenarios}} and \nameref{methods}).
(c)~Line plot (mean ± 95\% confidence band) showing the dependence of tRI on variations in variant fitness.
We increased the fitness parameter stepwise from~$0.98$ to $1.10$ for 1.7\% of mutations (variants with varied fitness), while keeping it fixed at $1.00$ for all other mutations (neutral variants).
(d)~Line plot (mean ± 95\% confidence band) showing the dependence of tRI on variations in sampling density.
We increased the sampling density stepwise from $10\%$ to $100\%$, with fitness fixed at $1.10$ for beneficial variants and at $1.00$ for neutral variants.
We note that tRI signals of neutral variants occurring in panels (c) and (d) arise from background topological noise in neutral evolution as in (a).
}
\label{fig:tri_simulated_scenarios_stats}
\end{figure}

To validate the reliability of the tRI as a robust indicator of positive selection, we employed simulations of genome evolution and statistical tests.
First, topological cycles can arise randomly in neutral evolution, causing some background noise in tRI signals.
However, from neutral evolutionary scenarios simulated with \mbox{SANTA-SIM}~\cite{Jariani2019}, we inferred that for SARS-CoV-2 this effect is very small and amounts to less than 5\% of topological/SNV cycles (\hyperref[fig:topological_noise]{Supplementary \autoref{fig:topological_noise}}) and less than 0.5\% of total tRI sum (\autoref{fig:tri_simulated_scenarios_stats} and \nameref{methods}).

Second, while consensus genomes obtained with current sequencing technologies have very high sequencing accuracy of almost 100\% \cite{Bull2020, Stoler2021, Bloom2023a}, sequencing errors can still affect phylogenetic analyses \cite{DeMaio2024, Turakhia2020, Steenwyk2023}.
To mitigate the impact of sequencing errors on tRI, we applied several filtering rules in the preprocessing of sequence alignments to ensure that only high-quality genomes were included in the topological recurrence analysis (\nameref{methods}).
Furthermore, because of the combinatorial structure of SNV cycles (each SNV must occur independently twice in SNV cycles; \autoref{fig:DefinitionOftRI}) it is unlikely that random nucleotide substitutions give rise to SNV cycles, and we therefore do not expect that stochastic sequencing errors drive tRI much.
To support this argument, we added simulated stochastic sequencing errors to a synthetic alignment generated under neutral evolution with SANTA-SIM, and to a large real SARS-CoV-2 whole genome alignment (\nameref{methods}).
We observed only a minimal tRI signal increase in the simulated alignment and none at all in the SARS-CoV-2 alignment (\autoref{fig:tri_simulated_scenarios_stats} and \hyperref[fig:sequencing_errors]{Supplementary \autoref{fig:sequencing_errors}}).

A Monte Carlo permutation test showed that the frequency distribution of raw tRI signals differs significantly from a uniform distribution (Kolmogorov–Smirnov test, $p<0.05$; \hyperref[fig:tri_frequency_distribution]{Supplementary \autoref{fig:tri_frequency_distribution}}).
We propose a conservative permutation-based tRI significance level to account for the effects of background noise and sequencing errors (\nameref{methods}).

On the other hand, systematic sequencing errors may confound results of the topological recurrence analysis---a phenomenon that is also present in phylogenetic inference \cite{DeMaio2024, Turakhia2020, Turakhia2021, Steenwyk2023, Hunt2024}.
To contain this effect, we recommend that in the downstream analysis of tRI signals, masking schemes \cite{TurakhiaData2021, DeMaio2024} are applied to exclude problematic sites.

Finally, we used synthetic evolutionary scenarios generated with SANTA-SIM to examine the sensitivity of the tRI for positive selection (\autoref{fig:tri_simulated_scenarios_stats}, \hyperref[fig:tri_simulated_scenarios]{Supplementary \autoref{fig:tri_simulated_scenarios}} and \nameref{methods}).
We found a very strong monotonic correlation (Spearman's~$\rho = 1.0$, $p < \num{1e-100}$) between fitness and tRI for variants with varied fitness.
The simulations moreover show that tRI is sensitive to beneficial variants, whereas deleterious variants produce almost no tRI signal.
Next, we quantified the dependence on sampling density of the effect size (Cohen's $d$) for the difference in tRI signals between beneficial and neutral variants.
We found a large effect size ($1.31 \leq d \leq 4.42$) for sampling densities ranging from $20\%$ to $100\%$, and a medium effect size ($d = 0.64$) for a sampling density of $10\%$.
Overall, these findings demonstrate that tRI is sufficiently sensitive to detect positive selection, and that this sensitivity is robust to variations in sampling density, topological noise and stochastic sequencing errors.

\begin{figure}[h!]
\centering
\includegraphics[width=\textwidth]{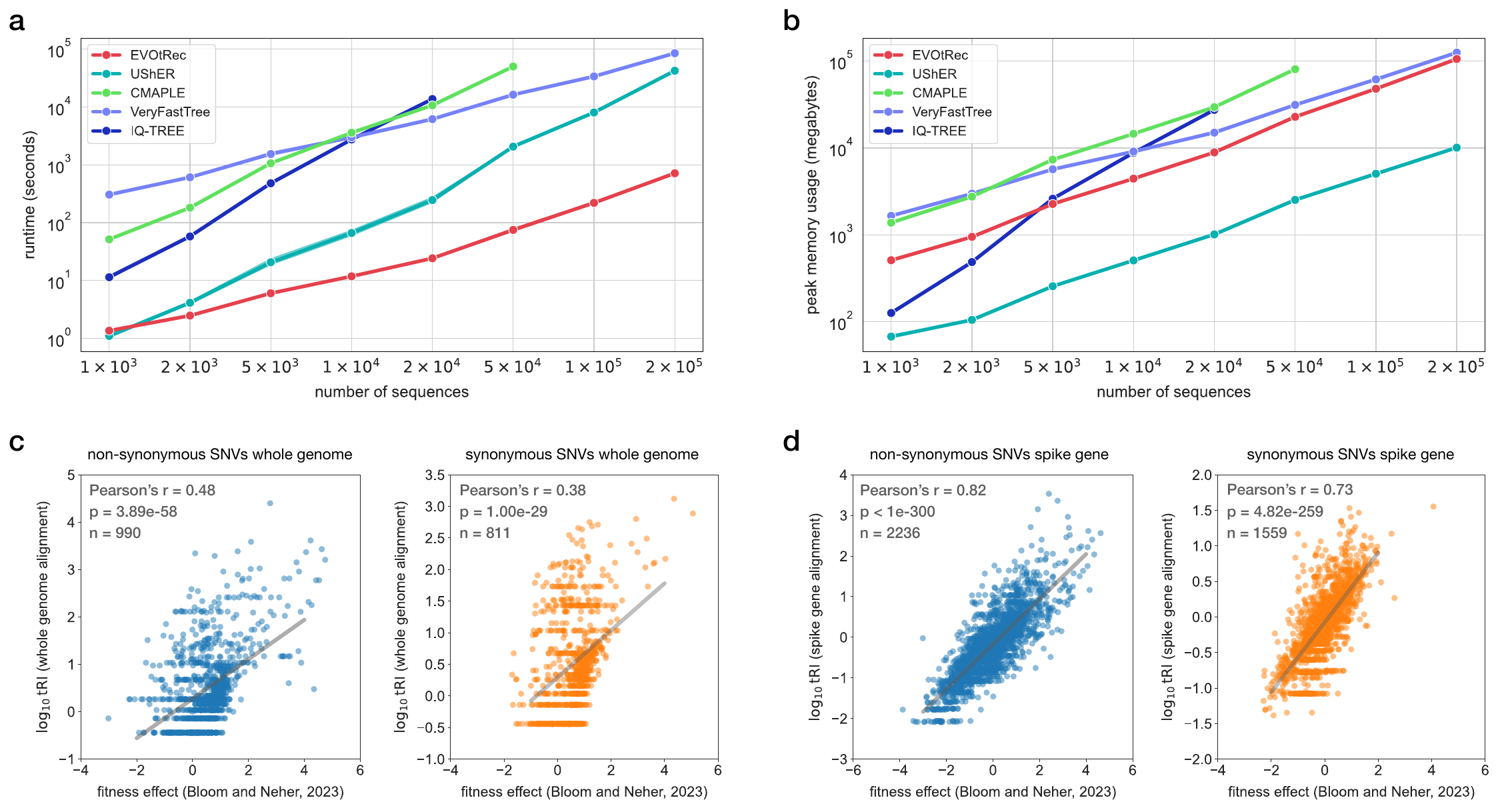}
\caption{\textbf{\sffamily{EVOtRec performs faster than phylogenetic tools with congruent results on fitness-increasing effects.}}
\textbf{(a, b)} Line plots (mean ± 95\% confidence band) showing runtime and peak memory usage (RAM) of EVOtRec vs.~the following phylogenetic inference tools: IQ-TREE \cite{Wong2025}, VeryFastTree \cite{Pieiro2024} and CMAPLE \cite{LyTrong2024} (de novo analysis), and UShER \cite{Turakhia2021} (phylogenetic sample placement).
For the benchmarking we used SARS-CoV-2 genome alignments with 1k up to 200k distinct sequences, obtained by subsampling the SARS-CoV-2 whole genome alignment (\hyperref[tab:tri_table_benchmarking]{Supplementary Table \ref{tab:tri_table_benchmarking}} and \nameref{methods}).
\textbf{(c, d)} Correlation between log-transformed positive tRI and phylogeny-based fitness effect estimates \cite{Bloom2023a} for SARS-CoV-2 SNVs on the whole genome (c) and the spike gene (d) (\nameref{methods}).
}
\label{fig:Benchmarking}
\end{figure}

\subsection{EVOtRec performs faster than phylogenetic tools with congruent results on fitness-increasing effects}

We compared EVOtRec with state-of-the-art phylogeny-based approaches for the assessment of fitness-increasing effects of genomic variants.
In a first step, we used SARS-CoV-2 whole genome alignments to benchmark the runtime and memory usage of EVOtRec against four tools for phylogenetic tree reconstruction: IQ-TREE \cite{Wong2025}, VeryFastTree \cite{Pieiro2024} and CMAPLE \cite{LyTrong2024} for de novo analysis, and UShER \cite{Turakhia2021} for phylogenetic sample placement (\autoref{fig:Benchmarking}, \hyperref[tab:tri_table_benchmarking]{Supplementary Table \ref{tab:tri_table_benchmarking}} and \nameref{methods}).
We found that for large alignments with more than 20k genomes, EVOtRec performed at least two orders of magnitude faster than IQ-TREE, \mbox{VeryFastTree} and CMAPLE, and more than one order of magnitude faster than UShER.
For example, EVOtRec completed the de novo analysis of the 200k genomes alignment in 11min 55s, yielding \textgreater100-fold speedup over the fastest de novo analysis tool (VeryFastTree, 23h 37min) and \textgreater50-fold speedup over the fastest sample placement tool (UShER, 11h 44min) (\hyperref[tab:tri_table_benchmarking]{Supplementary Table \ref{tab:tri_table_benchmarking}}).
Further, EVOtRec required more memory (RAM) than tools for phylogenetic sample placement (UShER), but less memory than tools for de novo phylogenetic inference (IQ-TREE, VeryFastTree and CMAPLE), with a peak memory usage of~106~GB for the 200k genomes alignment (UShER, 10~GB and VeryFastTree, 125~GB) (\hyperref[tab:tri_table_benchmarking]{Supplementary Table \ref{tab:tri_table_benchmarking}}).
EVOtRec gains computational efficiency by employing only short genetic distances, which make up less than 1\% of distances in large SARS-CoV-2 whole genome alignments.
This reduces the time (memory) consumption for the computation of persistence barcodes with Ripser \cite{Bauer2021} by~5~(3) orders of magnitude compared to the standard persistent homology computation (\hyperref[fig:ripser]{Supplementary \autoref{fig:ripser}}).

Next, we compared tRI signals with phylogenetic mutation count-based estimates of fitness effects.
Using simulated sequence alignments generated with SANTA-SIM, we first analyzed the Spearman correlation between raw tRI profiles and phylogenetic homoplasy counts computed with IQ-TREE and TreeTime \cite{Sagulenko2018} (\nameref{methods}).
We observed a moderate-to-strong correlation with $\rho = 0.59$ for beneficial variants, but no correlation for neutral variants ($\rho = 0.04$), with several neutral variants exhibiting no tRI signal but having positive homoplasy count (\hyperref[fig:simulated_tri_multiplicity]{Supplementary \autoref{fig:simulated_tri_multiplicity}}).

Moreover, mutation counts in the UShER phylogenetic tree have previously been used \cite{Bloom2023a} to estimate fitness effects of SARS-CoV-2 mutations.
We found a moderate-to-strong Pearson correlation between log-transformed tRI from our analysis and these fitness effect estimates \cite{BloomNeherData2024a} for the whole genome ($0.38 \leq r \leq 0.48$) and spike gene ($0.73 \leq r \leq 0.82$) (\autoref{fig:Benchmarking} and \nameref{methods}).
Divergence between tRI and phylogeny-based fitness effect estimates likely reflects the fact that tRI is sensitive to beneficial mutations, while fitness effect estimates have been reported to be better for estimating effects of deleterious or nearly neutral mutations in comparison with clade-growth methods \cite{Bloom2023a, Bloom2023b}.

These findings highlight that our approach, based on topological recurrence, differs from traditional recurrence analyses on global phylogenetic trees in that it analyzes the independent occurrence of mutations within topological cycles locally in the genomic dataset, without constructing a phylogenetic tree (\autoref{fig:DefinitionOftRI}).

\subsection{EVOtRec reveals the pandemic-scale dynamics of SARS-CoV-2 convergent evolution}

Convergent mutations play a key role in SARS-CoV-2 evolution and are indicative of potential adaptation \cite{vanDorp2020, Rochman2021, Zahradnik2022, Kistler2022, Markov2023}.
To demonstrate EVOtRec's ability to map the pandemic-scale dynamics of convergent evolution in time-resolved genomic datasets, we performed a time series topological recurrence analysis of SARS-CoV-2 genome data shared via GISAID, the global data science initiative \cite{Elbe2017}, collected between January 2020 and February 2024 and covering the entire COVID-19 pandemic.
The raw dataset contains \NumberOfGISAIDCoronaSequences whole genomes along with temporal information given by collection date.
We analyzed two high quality nucleotide sequence alignments which we compiled from this dataset, one containing whole genomes and another one containing spike genes (\hyperref[tab:supp_tri_whole_genome]{Supplementary Table \ref{tab:supp_tri_whole_genome}} and \nameref{methods}).
Restricting the analysis to the spike gene alignment resulted in stronger tRI signals, and provides a more detailed picture of convergence in the spike gene (\hyperref[fig:tri_region_subalignment]{Supplementary \autoref{fig:tri_region_subalignment}} and \nameref{methods}).

\begin{figure}[h!]
  \centering
  \includegraphics[width=\textwidth]{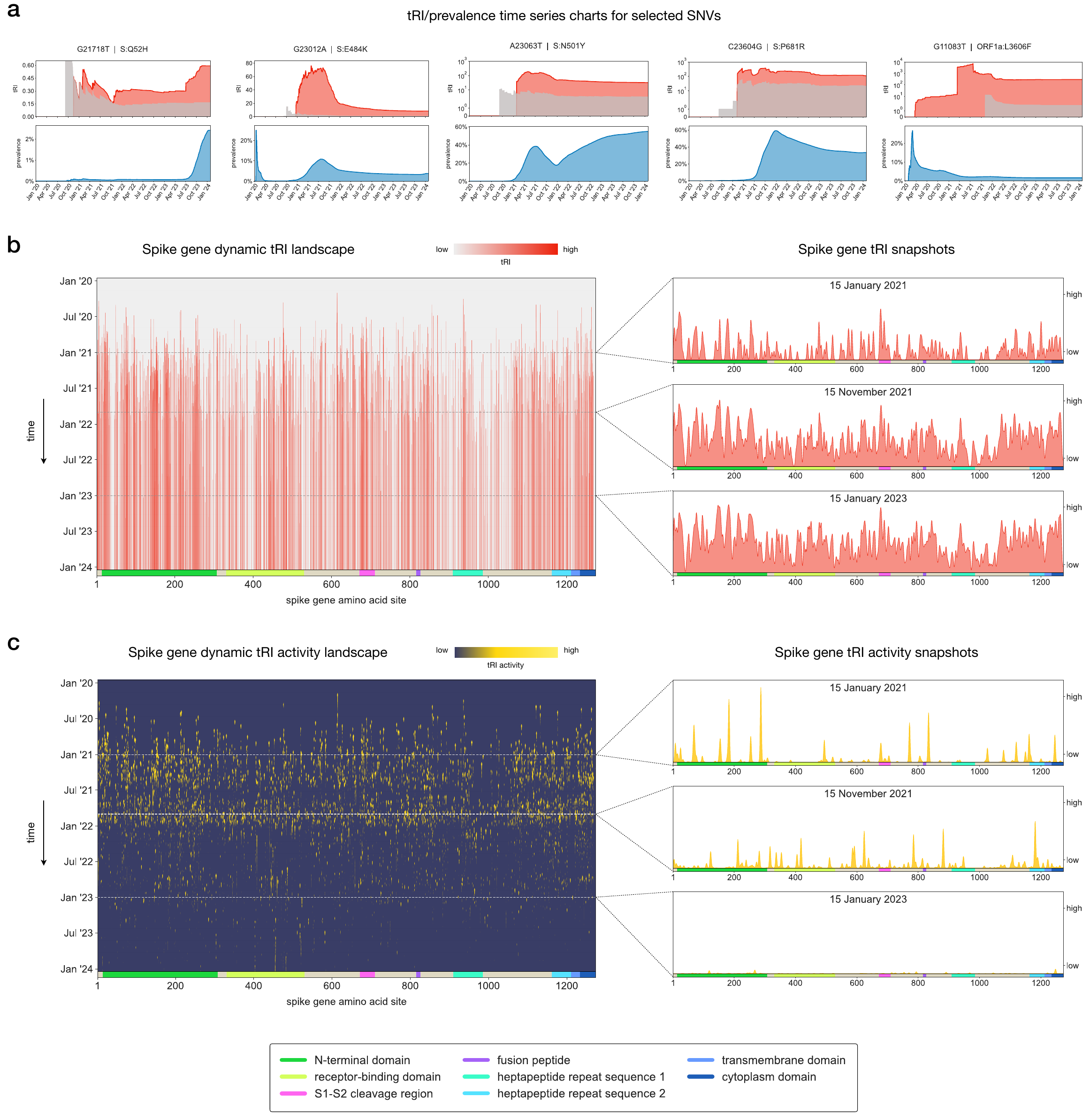}
  \caption{\textbf{\sffamily{EVOtRec reveals the pandemic-scale dynamics of SARS-CoV-2 convergent evolution.}}
We performed a time series topological recurrence analysis of the evolution of SARS-CoV-2 during the COVID-19 pandemic \cite{Markov2023} with EVOtRec, analyzing the whole genome and spike gene alignments based on data from GISAID \cite{Elbe2017} collected between January 2020 and February 2024 (\hyperref[tab:supp_tri_whole_genome]{Supplementary Table \ref{tab:supp_tri_whole_genome}} and \nameref{methods}).
\textbf{(a)} Time series charts showing tRI (red) and prevalence (blue) at daily resolution.
The area shaded in grey marks the tRI significance level (i.e.~the red tRI signal is significant whenever it is above the grey area).
Charts for the fitness-increasing spike amino acid changes Q52H \cite{Dadonaite2024a, Roemer2023}, E484K \cite{Collier2021, Liu2021a}, N501Y \cite{Liu2021c, Liu2021d, Martin2021} and P681R \cite{Liu2022} obtained from the spike gene alignment, and for the highly recurrent mutation L3606F \cite{Turakhia2020, DeMaio2024, Morel2020, vanDorp2020} on the ORF1a gene obtained from the whole genome alignment.
\textbf{(b, c)} Dynamic fitness landscapes mapping the spike gene convergent evolution throughout the entire COVID-19 pandemic from January 2020 until January 2024 (left), and snapshots (horizontal slices) showing fitness landscapes at fixed time points in the early, middle and late phase of the pandemic (right).
Dynamic tRI landscapes~(b) comprehensively capture time-dependent tRI signals across all amino acid sites on the spike gene.
Vertical line segments (red) indicate the strength of tRI signals over time at a specific site.
Dynamic tRI activity landscapes (c) capture changes in the intensity of convergent evolution by mapping positive growth rates of tRI signals in (b).
Vertical line segments (yellow) indicate periods of time during which tRI signals at a specific amino acid site were increasing.
}
    \label{fig:DynamicsOftRI}
\end{figure}

We used EVOtRec to study the convergent evolution of single nucleotide variations by means of \emph{tRI/prevalence time series charts} showing time-dependent tRI vs.~time-dependent prevalence at daily resolution (\nameref{methods}).
For example, for the beneficial spike gene amino acid changes Q52H \cite{Dadonaite2024a, Roemer2023}, E484K \cite{Collier2021, Liu2021a}, N501Y \cite{Liu2021c, Liu2021d, Martin2021} and P681R \cite{Liu2022} we observed an increase in tRI over a certain period of time, which is accompanied, or followed, by an increase also in prevalence (\autoref{fig:DynamicsOftRI}).
These charts can be used for the real-time tracking of convergently evolving genomic variants and to identify time intervals in which a specific mutation potentially conferred a selective advantage.
We found that on the last day of each time series (29/30 January 2024), tRI and prevalence were moderately correlated with Pearson's $r = 0.24$ for the whole genome alignment ($r = 0.25$ for the spike gene alignment) (\hyperref[fig:tri_prevalence]{Supplementary \autoref{fig:tri_prevalence}}).
This is consistent with positive selection acting as a confounding factor, where evolutionary advantages lead to an increase in both tRI and prevalence.
For some mutations we observed a different pattern---a typical example is the highly recurrent mutation L3606F \cite{Turakhia2020, DeMaio2024, Morel2020, vanDorp2020} on the ORF1a gene, for which tRI and prevalence showed diverging trends (\autoref{fig:DynamicsOftRI}), suggesting that this mutation is neutral or deleterious.

We recovered the known convergently evolving spike gene amino acid sites R346, K444, L452, F456, A475, N460 and F486.
Amino acid changes at these sites have been shown to recurrently occur in several Omicron sub-lineages \cite{Cao2022, Ito2023, Jian2024}, and we detected significant tRI signal at each of these sites after the emergence of the Omicron variant in late 2021 (\hyperref[fig:tri_convergent_mutations]{Supplementary \autoref{fig:tri_convergent_mutations}}).

While tRI/prevalence time series charts map the convergent evolution of individual SNVs, a more comprehensive picture can be obtained by simultaneously comparing time-dependent tRI signals across all amino acid sites on the genome.
For this, we integrated EVOtRec tRI time series data into a \emph{dynamic tRI landscape} which captures tRI signals across the entire spike gene for the period from January 2020 until February 2024 at daily resolution (\autoref{fig:DynamicsOftRI} and \nameref{methods}).
Landscapes of tRI signals at specific time points can then be extracted by taking \emph{snapshots} (horizontal slices) in the dynamic tRI landscape.
To localize hotspots of convergent evolution both in time and genome position, we further integrated EVOtRec time series data into a \emph{dynamic tRI activity landscape} that visualizes positive tRI growth rates across all amino acid sites on the spike gene in a two-dimensional heatmap (\autoref{fig:DynamicsOftRI} and \nameref{methods}).
Applying Gaussian smoothing can help to reduce image detail and enhance hotspots of convergent evolution (\autoref{fig:Adaptation}).
In summary, landscapes of tRI signals enable the rapid and convenient identification of genomic regions shaped by convergent evolution over time.
The interpretation should be supported at the level of individual SNVs by more detailed tRI/prevalence charts and, if available, additional experimental data.

\subsection{Topological recurrence correlates with adaptation}

\begin{figure}[h!]
  \centering
  \includegraphics[width=16cm]{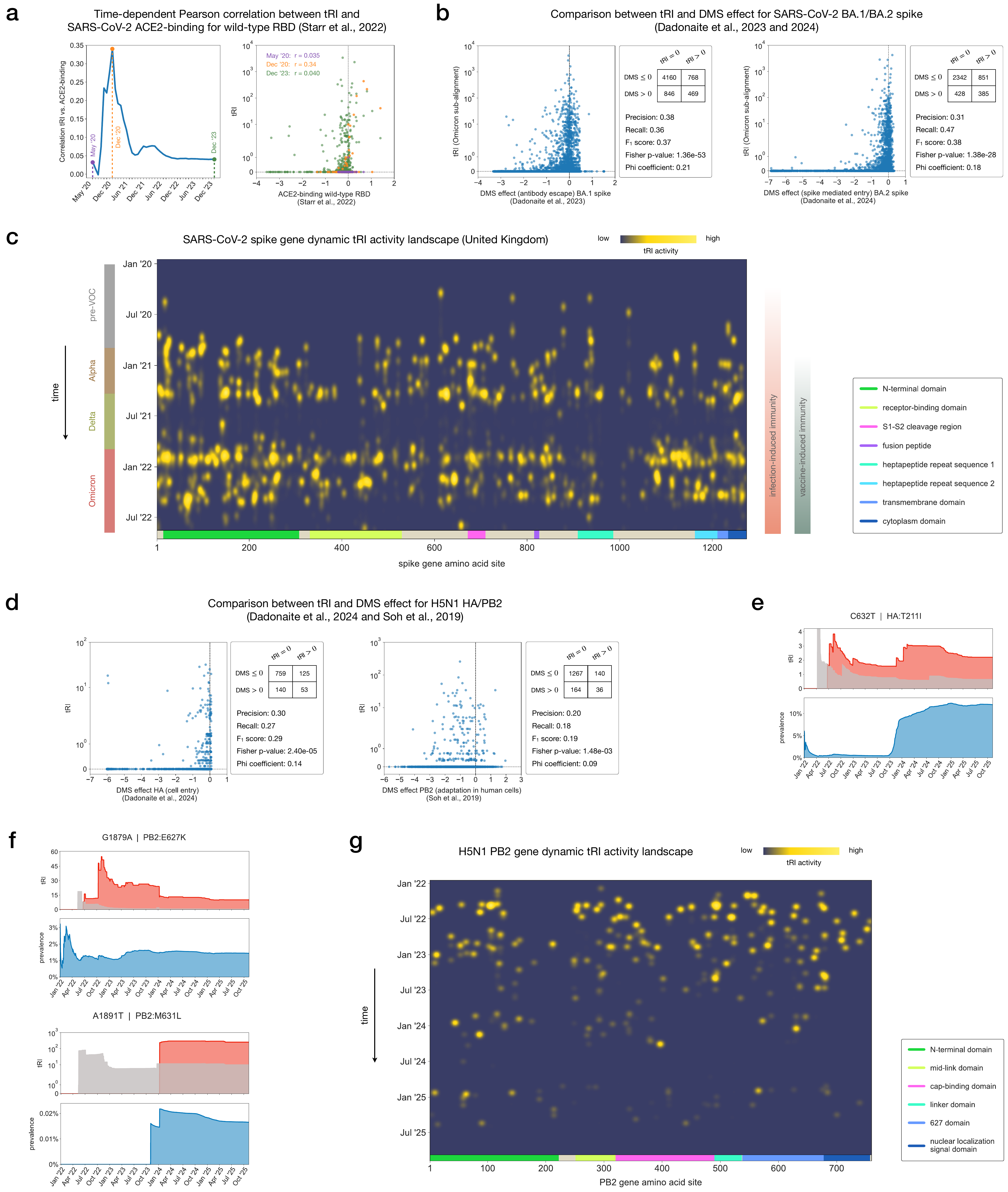}
  \caption{\textbf{\sffamily{Topological recurrence correlates with adaptation.}}
\textbf{(a)} Time-dependent Pearson correlation between tRI and ACE2-binding affinity for amino acid changes in the \mbox{SARS-CoV-2} wild-type RBD (\nameref{methods}).
Experimental data taken from yeast display deep mutational scanning (DMS) measurements \cite{Starr2022a}.
\textbf{(b)} Comparison between tRI and published pseudovirus DMS measurements (antibody escape and cell entry) \cite{Dadonaite2023, Dadonaite2024a} of amino acid changes in the SARS-CoV-2 Omicron spike gene alignment for the BA.1 and BA.2 sub-lineages.
Statistical analyses are based on contingency tables for binarized tRI/DMS effects (\nameref{methods}).
\textbf{(c)} SARS-CoV-2 spike gene dynamic tRI activity landscape for the United Kingdom covering the period from January~2020 until August~2022 (\nameref{methods}).
Hotspots of tRI activity reflect changes in genomic background (emergence of dominant genomic variants) and in host immunity status (infection and vaccination) \cite{Carabelli2023}.
\textbf{(d)} Comparison between tRI and published pseudovirus DMS measurements (cell entry and adaptation in human cells) \cite{Dadonaite2024b, Soh2019} of amino acid changes in the H5N1 HA and PB2 gene alignments.
Statistical analyses as in (b).
\textbf{(e-f)} Time series charts showing tRI (red; area shaded in grey marks tRI significance level) and prevalence (blue) at daily resolution for the mammalian-adaptive mutations HA:T211I (T199I in H3 numbering) \cite{Good2024}~(e) and PB2:E627K \cite{Carrique2020, Peacock2023} and PB2:M631L \cite{Dholakia2026} (f).
\textbf{(g)} H5N1 PB2 gene dynamic tRI activity landscape covering the period from January~2021 until November~2025 (\nameref{methods}).
Hotspots of tRI activity reflect changes in genomic background (emergence of genotypes B3.13 in late 2023/early 2024 and D1.1 in late 2024/early 2025).
}
  \label{fig:Adaptation}
\end{figure}

For the biological validation of our method, we compared EVOtRec tRI signals with published experimental and epidemiological data characterizing adaptation for three different viruses: SARS-CoV-2, influenza A subtype H5N1 and HIV-1.
First, we demonstrated EVOtRec's ability to capture the convergent evolution in pandemic-scale SARS-CoV-2 data by comparing spike gene tRI signals with published deep mutational scanning (DMS) measurements for the wild-type and the Omicron BA.1, BA.2 and XBB.1.5 sub-lineages.
We found that the Pearson correlation between time-dependent tRI obtained from the spike gene alignment and ACE2-binding affinity for the wild-type receptor-binding domain (RBD) \cite{Starr2022a} was rising during the first 10 months of the COVID-19 pandemic (\autoref{fig:Adaptation}) when the wild type was the predominant circulating variant in an immunologically naive host population \cite{Carabelli2023}, before it began to decline again as immune escape became an increasingly important evolutionary factor \cite{Martin2021, Carabelli2023} (\nameref{methods}).
We moreover investigated the performance of tRI in identifying fitness-increasing amino acid changes (i.e.~positive DMS effect) for the BA.1, BA.2 and XBB.1.5 sub-lineages.
We found a statistically significant association between binarized tRI ($\textrm{tRI} > 0$ vs.~$\textrm{tRI} = 0$) obtained from the Omicron spike gene alignment and binarized DMS effect ($\textrm{DMS effect} > 0$ vs.~$\textrm{DMS effect} \le 0$) for antibody escape in BA.1 \cite{Dadonaite2023} and spike-mediated entry in BA.2 \cite{Dadonaite2024a} (Fisher's exact test, $p<0.01$; \autoref{fig:Adaptation} and \nameref{methods}).
For XBB.1.5, a logistic model predicting positive tRI signals as a function of the full three-dimensional XBB.1.5 DMS data \cite{Dadonaite2024a} suggests a correlation with ACE2-binding ($\beta_1 = 0.53$), spike mediated entry ($\beta_2 = 1.44$) and human sera escape ($\beta_3 = 1.02$) (\hyperref[fig:stats_tri_fitness_dms]{Supplementary \autoref{fig:stats_tri_fitness_dms}} and \nameref{methods}).
Findings from this logistic model are consistent with results from a similar phylogeny-based ordinary least squares regression analysis using published fitness effect estimates \cite{Bloom2023a, Haddox2025} (\hyperref[fig:stats_tri_fitness_dms]{Supplementary \autoref{fig:stats_tri_fitness_dms}}).
The strength of correlations between tRI and DMS measurements is likely limited by differences in mutational and selective factors in virus transmission in the wild versus selective experiments in the laboratory.
Mutational factors include host RNA editing processes \cite{DiGiorgio2020, Graudenzi2021, Simmonds2020, Ratcliff2021, Zhao2022a, Bloom2023b} and hyper-mutable sites \cite{DeMaio2024, Simmonds2020}.
Selective factors include transmissability \cite{Markov2023}, as well as epistatic interactions across the whole genome or varying genomic backgrounds \cite{Moulana2022, Moulana2023, Javanmardi2022, Bloom2023b}.
Aside from biological differences, systematic sequencing errors can also cause apparent genomic recurrence \cite{DeMaio2024, Turakhia2020}, increasing tRI at some sites. 

We further used these experimental DMS data to compare the performance of tRI with phylogeny-based fitness effect estimates \cite{Bloom2023a, Haddox2025} in identifying fitness-increasing amino acid changes (i.e.~positive DMS effect) for three different SARS-CoV-2 phenotypes (\hyperref[fig:stats_tri_fitness_dms]{Supplementary \autoref{fig:stats_tri_fitness_dms}}).
We found that both methods yield comparable $F_{1}$ score and precision, while the tRI tends to exhibit higher recall.
Notably, our unbiased, organism- and site-agnostic approach achieves predictive performance on par with the phylogeny-based fitness effect estimates in \cite{Haddox2025} which invoke SARS-CoV-2 site-specific mutation rates and sequence context.

Moreover, we observed that at least 84\% of all amino acid changes that had fixed in the Alpha, Beta, Gamma, Delta or Omicron variants also exhibited significant tRI, while among all amino acid changes only 15\% had this property (\hyperref[fig:tri_frequency_distribution]{Supplementary \autoref{fig:tri_frequency_distribution}}).

To assess EVOtRec's ability to map the time-dependence of convergent evolution in the context of changing genomic backgrounds, we investigated to what extent the dynamic tRI landscape of the SARS-CoV-2 spike gene was modulated by an increase in host immunity and the emergence of dominant virus variants over the course of the COVID-19 pandemic in the United Kingdom \cite{Carabelli2023, Markov2023, Vhringer2021, duPlessis2021} between January~2020 and  August~2022 (\autoref{fig:Adaptation}, \hyperref[tab:supp_tri_uk_spike_gene]{Supplementary Table~\ref{tab:supp_tri_uk_spike_gene}} and \nameref{methods}).
We observed a significant rise in tRI activity in the second half of~2020, with topological signals accumulating in the S1-S2 cleavage region, which is a determinant of SARS-CoV-2 transmissibility \cite{Huang2020, Johnson2021, Peacock2021, Carabelli2023}, in regions in the RBD which are known to enhance ACE2-binding, such as for example site N501 \cite{Liu2021c, Liu2021d, Zahradnik2021a, Martin2021}, and in the N-terminal domain which allosterically modulates S1-S2 cleavage and the conformation of the RBD for ACE2-binding \cite{Meng2022, Dadonaite2024a}.
This is in line with the observation that adaptation during the first year of the pandemic was mainly enhancing infectiousness and transmissibility in an immunologically naive host population, as seen in the Alpha and Delta variants \cite{Volz2021, Carabelli2023}.
Immune escape became an evolutionary factor early in the year 2021, when infection-induced population immunity was on the rise on the Alpha and Delta genomic backgrounds \cite{Martin2021, Zahradnik2022, Carabelli2023}.
Correspondingly, we recorded a further increase in tRI signals in the N-terminal domain and the RBD, which contain epitopes for antibody binding \cite{Harvey2021, Huang2020, Piccoli2020, Dejnirattisai2021, Chen2022, Klinakis2021} such as site E484 \cite{Collier2021, Liu2021a}, and in the S1-S2 cleavage region \cite{Meng2022}, as for example at site P681 \cite{Liu2022} (\autoref{fig:Adaptation}).
After a longer period with almost no tRI activity, we observed another significant rise in activity towards the end of the year 2021, with distinct hotspots in the N-terminal domain, the RBD, the S1-S2 cleavage region, the fusion peptide and other regions in the S2 subunit, which play a key role in immune escape \cite{Harvey2021, Piccoli2020, Barnes2020, Dejnirattisai2021, Chen2022, Lei2024, Dadonaite2024a, Jian2023, Cao2022, Roemer2023b, Javanmardi2022} (\autoref{fig:Adaptation}).
This is congruent with the observation that during this later phase of the pandemic, immune evasion became the dominant factor of positive selection in the presence of widespread population immunity, as seen in the convergent evolution of the Omicron variant and its sub-lineages \cite{Carabelli2023, Cao2022, Willett2022, Roemer2023b, Ito2023, Jian2024}.

These findings indicate that dynamic tRI activity landscapes can capture shifts in variant fitness resulting from changes in genomic background and interactions with genetic changes at other sites in the genome \cite{Martin2021, Zahradnik2022, Johnson2023} (\autoref{fig:Adaptation}).
It is a particular feature of our method that, since genomes in SNV cycles share the same genomic background (\autoref{fig:DefinitionOftRI}), tRI automatically keeps track of shifting genomic backgrounds without the need to analyze sub-alignments corresponding to different phylogenetic clades (\hyperref[fig:tri_region_subalignment]{Supplementary \autoref{fig:tri_region_subalignment}}).

\begin{table}[h!]
  \centering
  \includegraphics[width=14cm]{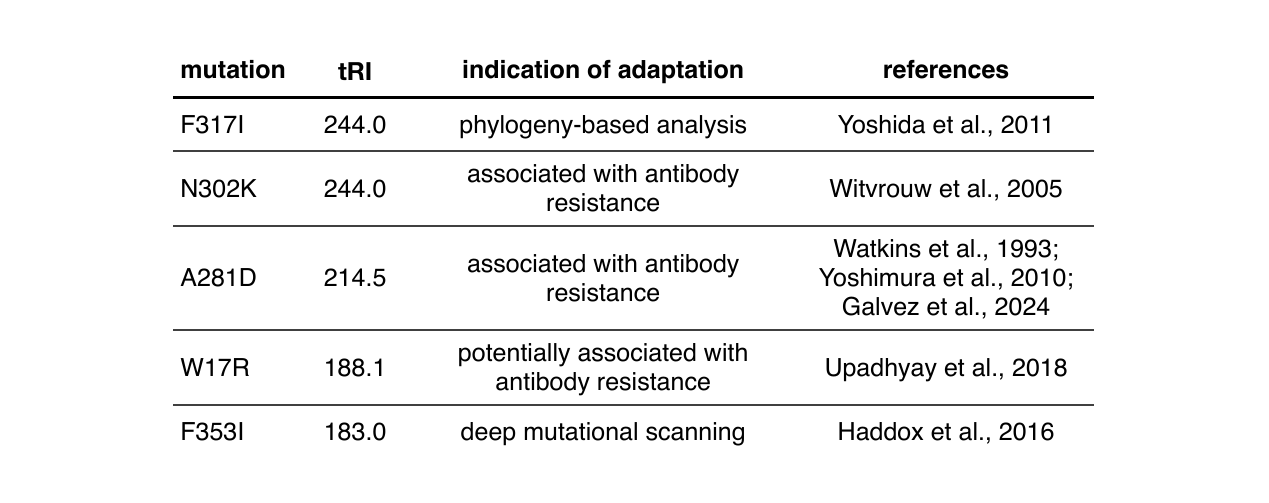}
  \caption{\textbf{\sffamily{Topological recurrence correlates with adaptation.}}
  Comparison of tRI with published data on fitness increasing mutations in HIV-1.
The table shows the top five amino acid changes on the HIV-1 envelope gene (Env) with strongest tRI signal.
These five substitutions have been associated with (potential) adaptation in the literature (see \hyperref[tab:tri_table_hiv]{Supplementary \autoref{tab:tri_table_hiv}} for a more comprehensive list).
The tRI data was generated with EVOtRec from the Env gene alignment (\hyperref[tab:supp_tri_hiv_env]{Supplementary Table~\ref{tab:supp_tri_hiv_env}} and \nameref{methods}), and the indication of adaptation is taken from \cite{Yoshida2011, Witvrouw2005, Watkins1993, Yoshimura2010, Galvez2024, Upadhyay2018, Haddox2016}.
}
  \label{tab:hiv-adaptation}
\end{table}

Finally, we examined EVOtRec's ability to identify fitness-increasing genomic variants in the evolution of influenza A subtype H5N1 and HIV-1.
Since the sampling rate in the corresponding alignments is lower than in SARS-CoV-2, we augmented the data using ancestral sequence reconstruction (\hyperref[fig:data_augmentation]{Supplementary \autoref{fig:data_augmentation}} and \nameref{methods}).
For H5N1, we performed time series topological recurrence analyses of the hemagglutinin gene (HA) and the polymerase basic~2 gene (PB2) alignments belonging to the currently circulating highly pathogenic avian influenza H5N1 strain~2.3.4.4b (\hyperref[tab:supp_tri_influenza_ha]{Supplementary Table~\ref{tab:supp_tri_influenza_ha}} and \nameref{methods}).
Using tRI data from the last day of each time series (7/8 November 2025), we found a statistically significant association between binarized tRI ($\textrm{tRI} > 0$ vs.~$\textrm{tRI} = 0$) and binarized DMS effect ($\textrm{DMS effect} > 0$ vs.~$\textrm{DMS effect} \le 0$) for pseudovirus entry into 293T cells for the HA gene \cite{Dadonaite2024b} and viral adaptation in human A549 cells for the PB2 gene \cite{Soh2019} (Fisher's exact test, $p<0.01$; \autoref{fig:Adaptation} and \nameref{methods}).
The observed association between tRI and DMS measurements is less strong for H5N1 than for SARS-CoV-2, which may be explained by the fact that the percentage of mutations in the HA and PB2 alignments among all mutations that have been tested in the DMS experiments is lower than in \mbox{SARS-CoV-2} (\hyperref[fig:stats_mutations_in_dms]{Supplementary \autoref{fig:stats_mutations_in_dms}}).
We detected statistically significant tRI signals for key signatures of mammalian adaptation \cite{Mostafa2024, Peacock2024, Capelastegui2025, Nguyen2025}, including the amino acid substitutions T211I (T199I in H3 numbering) \cite{Good2024} in the HA gene, and E249G \cite{Yamaji2015}, A588T \cite{Fan2014}, E627K \cite{Carrique2020, Peacock2023}, M631L \cite{Gu2024, Dholakia2026}, D701N \cite{Gao2009, Peacock2023} and  D740N \cite{Dholakia2026} in the PB2 gene (\autoref{fig:Adaptation} and \hyperref[fig:tri_convergent_mutations]{Supplementary \autoref{fig:tri_convergent_mutations}}).
Moreover, the dynamic tRI activity landscape of the PB2 gene (\autoref{fig:Adaptation} and \nameref{methods}) reveals patterns of topological signals likely associated with changes in genomic background following spillover events from wild birds to dairy cattle in North America \cite{Nguyen2025, Dholakia2026, Krammer2025}.
For example, we observed an increase in tRI for the key adaptive mutations M631L \cite{Dholakia2026} in the B3.13 genotype in late~2023/early~2024, and D701N~\cite{Pekar2025} in the D1.1 genotype in early~2025.
We further report HA and PB2 amino acid changes currently under convergent evolution (\hyperref[fig:h5n1_convergent_mutations]{Supplementary \autoref{fig:h5n1_convergent_mutations}}).

For HIV-1, we performed the tRI analysis for the envelope gene (Env), which is essential for host cell binding and entry, and has the capability to escape host immunity through its rapid evolution \cite{Yoshida2011, Arrildt2012} (\hyperref[tab:supp_tri_hiv_env]{Supplementary Table~\ref{tab:supp_tri_hiv_env}} and \nameref{methods}).
We found that among the top~40 amino acid changes on the Env gene with strongest tRI signals, at least 70\% have been associated with potential positive selection in the literature (\autoref{tab:hiv-adaptation} and \hyperref[tab:tri_table_hiv]{Supplementary \autoref{tab:tri_table_hiv}}).

\subsection{Discussion}

In this study, we introduce a method from Topological Data Analysis that can rapidly identify the presence of recurrent genomic variants in pandemic-scale sequence alignments.
Our approach uses persistent homology to quantify convergent evolution in terms of global topological patterns in genomic datasets, without relying on the challenging prior reconstruction of an accurate phylogenetic tree.
We implemented the method in the EVOtRec pipeline, which in benchmark tests performed orders of magnitude faster than state-of-the-art tree inference tools like IQ-TREE \cite{Wong2025}, VeryFastTree \cite{Pieiro2024}, UShER \cite{Turakhia2021} and CMAPLE \cite{LyTrong2024}, while results were congruent with phylogeny-based fitness effect estimates \cite{Bloom2023a, Haddox2025}.
The pipeline is distance-based and organism-agnostic, and enables an unbiased and scalable analysis of millions of genomes, which facilitates leveraging the ever-increasing wealth of genomic sequencing data.

Our method is designed for the rapid analysis of the dynamics of convergent evolution in very large time-resolved genomic datasets.
To demonstrate its utility and versatility, we applied the EVOtRec pipeline to pandemic-scale viral genome datasets from SARS-CoV-2, influenza A subtype H5N1 and HIV-1, representing three different virus families.
The SARS-CoV-2 dataset we analyzed contained more than $14$ million genomes, covering the entire COVID-19 pandemic.
We showed that EVOtRec’s speed and scalability enable the real-time monitoring of emerging potentially adaptive mutations in ongoing evolutionary processes over time.
In addition, the method reveals hotspots of convergent evolution across a two-dimensional dynamic fitness landscape by mapping genomic recurrence simultaneously in both time and genome position.
The method is also well-suited to the retrospective analysis of convergent evolution in historical genomic data.
We validated our approach with simulated data and published experimental data (deep mutational scanning) across the three viruses, confirming that EVOtRec can robustly identify fitness-increasing genomic variants.
Notably, we demonstrated that our method matches or exceeds the predictive power of the most biologically-informed phylogeny-based approaches \cite{Bloom2023a, Haddox2025}.

Care should be taken in the interpretation of EVOtRec tRI signals. Our method resolves homoplasies only at small genetic distance scales, does not distinguish between forward and backward mutations, and does in its present form not include insertions and deletions.
While we applied several filtering rules in the preprocessing of sequence data and our analyses indicate that EVOtRec is robust to stochastic sequencing errors, systematic sequencing errors may still confound tRI signals.
Moreover, we observed that mutations identified as topologically recurrent need not necessarily be adaptive (e.g.~at hyper-mutable sites).
We therefore recommend that in the downstream analysis, additional biological information, such as masking schemes, is used to identify problematic sites.
Finally, the computational performance of EVOtRec may depend on the sampling rate, the ordering of sequences in the alignment and the topological structure of the genomic dataset.

Future developments in this method for viral comparative genomics could include normalizing the tRI by more detailed or site-specific mutational spectra, and investigating the utility of our approach for finding sites under purifying selection or subject to epistatic interactions.
The method can motivate and guide experimental studies by reducing the biologically relevant combinatorial complexity of possible genetic variants to explore in the laboratory.
Promising applications beyond viral genomics could support studies of bacterial evolution and somatic mutation processes in cancer.
Analyzing evolution over greater genomic distances will come with new challenges, but the general method presented here is highly adaptable to new data types including amino acid alignments from bacterial genes or somatic mutation profiles in cancer genomes.
The growing size of comparative genomics demands innovative approaches, and we propose Topological Data Analysis as a method that opens the way for new data-driven insights into molecular evolution at large scale.

\section{Methods}
\label{methods}

\subsection{Topological Data Analysis of genome evolution}

Topological Data Analysis \cite{Carlsson2009} originated from the mathematical field of \emph{algebraic topology} and studies the shape of datasets by extracting topological structures and patterns. Such topological structures have associated dimensions: structures of dimension zero can be thought of as the connected components, and structures of dimension one are essentially the loops, or topological cycles, of the dataset. Structures of higher dimensions can also be defined, but are also more difficult to interpret.
Here we are interested in reticulate evolutionary processes, thus we choose to focus on topological structures in dimension one, since topological cycles can be interpreted as signals of divergence from phylogenetic trees (\hyperref[fig:DefinitionOftRI]{\autoref{fig:DefinitionOftRI}a}).

Datasets often come as point clouds: in our setting, each point corresponds to a virus genome sample, and lies in a high-dimensional space where each nucleotide of the genome is a dimension~\cite{Rabadan2019}. A common way to extract the phylogenetic network from this point cloud simply amounts to connecting samples as soon as their genetic distance is less than a given threshold ${r>0}$. This results in a (neighborhood) graph, whose set of cycles provides candidates for the topological structures in dimension one of the true underlying network.
However, a main limitation of this approach comes from the fact that relevant topological structure typically appears at multiple scales.

The most common way to handle this issue in Topological Data Analysis is to actually compute and track the cycles for all possible values of~$r$ ranging from~$0$ to~$+\infty$. As~$r$ increases, some cycles can appear, and some already existing cycles can disappear, or get filled in.
The whole point of Topological Data Analysis is to record, for each cycle, its radius of appearance, or birth time, and radius of disappearance, or death time. This is called the \emph{persistent homology} of the point cloud \cite{Carlsson2009, Rabadan2019}. The construction, based on a scale parameter~$r$, can be summarized as follows (\hyperref[fig:ripser]{Supplementary \autoref{fig:ripser}a}).
The input is a distance matrix describing the dataset, considered as a finite metric space. First, consider the \emph{geometric graph} at scale~$r$, whose vertices are the data points, with any two points connected by an edge whenever their distance is at most~$r$. Generalizing this construction, the \emph{Vietoris--Rips complex} at scale~$r$ connects any subset of the data points with a simplex (an edge, a triangle, a tetrahedron, or a higher-dimensional generalization thereof) whenever all pairwise distances in the subset are at most~$r$. A Vietoris--Rips complex is a particular type of \emph{simplicial complex}, a higher-dimensional generalization of graphs which is of crucial interest in algebraic topology, in particular in homology theory. The family of Vietoris--Rips complexes for all parameters~$r$ is called the \emph{Vietoris--Rips filtration}. It provides a multiscale method to extract cycles of various sizes, and to encode them in a so-called \emph{persistence barcode}: each bar, or interval, in this barcode corresponds to a \emph{topological cycle} representing a topological feature (a reticulate evolutionary process in the present case), and the bar endpoints correspond to its radii of birth and death (the maximum genetic distance between consecutive samples forming the cycle, and, roughly, the maximum pairwise genetic distance between samples forming the cycle).

Each bar indicates the presence of a reticulate event, implying that the evolutionary history cannot be fully explained in terms of a single phylogenetic tree \cite{Chan2013, Camara2016, Rabadan2019}.
The mathematical background of this phenomenon is a classical theorem due to Rips, which asserts that trees have trivial persistent Vietoris--Rips homology~\cite{Gromov1987, Chan2013}.
The corresponding cycle in the associated reticulate phylogeny can then be localized in the sequence alignment by tracing it back to the isolates that constitute the reticulate event.
Moreover, the length of the bar represents the cycle size. In our case, this corresponds to the length of the reticulate evolutionary process, which allows to distinguish, for instance, between homoplasies and recombinations (\hyperref[fig:DefinitionOftRI]{\autoref{fig:DefinitionOftRI}b}).

\subsection{The EVOtRec pipeline}

The EVOtRec pipeline uses persistent homology to quantify convergent evolution in genomic datasets.
For time-resolved datasets, EVOtRec can map the time-dependence of these signals.
A basic version of the EVOtRec pipeline that computes raw tRI is implemented in the Python script evotrec.py (\url{https://github.com/ottamj/evotrec}).
We applied the pipeline to SARS-CoV-2, H5N1 and HIV-1 datasets, proceeding in three main steps (\hyperref[fig:DefinitionOftRI]{\autoref{fig:DefinitionOftRI}d}).
We first outline these steps and provide more details in subsequent sections.

\medskip

\noindent \textbf{1. Preprocessing.}
We downloaded and prepared a sequence alignment with a fixed reference sequence.
We used Hammingdist \cite{Keegan2024} to compute genetic distances in the sequence alignment.
For the time series analysis, we further applied the MuRiT algorithm \cite{Bleher2022} to rescale genetic distances depending on temporal information.
To improve efficiency, genetic distances were stored in a specifically designed sparse matrix format which keeps track of short genetic distances only.

\medskip

\noindent \textbf{2. Topological Data Analysis.}
We used Ripser \cite{Bauer2021} to compute the persistence barcode of the genetic distance matrix and to localize topological cycles in the genomic dataset.
Ripser outputs a list of topological cycles that correspond to the bars in the barcode.
The persistent homology computation is compatible with the EVOtRec sparse distance matrix format and was restricted to topological cycles in which all edges have short length (with respect to genetic distance).

\medskip

\noindent \textbf{3. Topological recurrence analysis.}
We performed a topological recurrence analysis on the Ripser output.
SNV cycles were extracted and evaluated for the computation of the raw topological recurrence index (raw tRI).
Time points of topological signals were recovered from the time-rescaled length of edges in SNV cycles, and corrections were applied to the raw tRI to obtain final tRI values.
Subsequently, in the downstream analysis time series tRI data were visualized using tRI charts and two-dimensional dynamic tRI landscapes.

\subsection{Acquisition and preparation of sequence alignments}

\subsubsection{Coronavirus}

We used SARS-CoV-2 genome data provided by the GISAID EpiCoV Database \cite{Elbe2017} and accessible at \url{https://doi.org/10.55876/gis8.240314pc} (alignment msa\_0222.fasta downloaded on 24~February~2024).
This alignment comprises \NumberOfGISAIDCoronaSequences SARS-CoV-2 nucleotide sequences aligned to the reference sequence hCoV-19/Wuhan/WIV04/2019.
We generated four sub-alignments by applying the following operations:
\begin{enumerate}[label=(\arabic*), topsep=1ex, itemsep=0.1ex, parsep=0.1ex]
  \item Sequences were truncated to specific genomic regions.
  \item High-quality genomes (complete, no ambiguous nucleotide sites, complete collection date) were selected.
  \item Only sequences that occurred in at least three identical copies in the alignment were retained, and among identical sequences, only the earliest (by collection date) was retained.
  \item Sequences in the alignment were arranged in time-reversed order (by collection date).
\end{enumerate}

\noindent\textbf{Whole genome alignment:} \NumberOfWholeGenomeSequences distinct whole genomes; length: \LengthOfWholeGenomeSequences nt; country: any; lineage: any; collection date: between December~2019 and February~2024.

\vspace{1mm}
\noindent\textbf{Spike gene alignment:} \NumberOfSpikeGeneSequences distinct spike genes; length: \LengthOfSpikeGeneSequences nt; country: any; lineage: any; collection date: between December~2019 and February~2024.

\vspace{1mm}
\noindent\textbf{Omicron spike gene alignment:} \NumberOfOmicronSpikeGeneSequences distinct spike genes; length: \LengthOfOmicronSpikeGeneSequences nt; country: any; lineages: Omicron BA.1, BA.2 and XBB.1.5; collection date: between November~2021 and February~2024.

\vspace{1mm}
\noindent\textbf{United Kingdom spike gene alignment:} \NumberOfUKSpikeGeneSequences distinct spike genes; length: \LengthOfUKSpikeGeneSequences nt; country: United Kingdom; lineages: any; collection date: between December~2019 and August~2022.

\vspace{1mm}
\noindent For more details see Supplementary Information Section~1.1.

\subsubsection{Avian influenza virus}

We used two avian influenza A subtype H5N1 (strain 2.3.4.4b) genome datasets, one for the HA gene and another one for the PB2 gene, provided by the GISAID EpiFlu Database~\cite{Elbe2017} and accessible at \url{https://doi.org/10.55876/gis8.251116eh} (HA gene) and \url{https://doi.org/10.55876/gis8.251116bc} (PB2 gene).
We downloaded all available nucleotide sequences (complete, with complete collection date not before 30~December~2021) as of 16~November~2025.
The HA dataset comprises \NumberOfGISAIDInfluenzaHASequences sequences and the PB2 dataset comprises \NumberOfGISAIDInfluenzaPBtwoSequences sequences.
We aligned each of these datasets to the reference sequence A/American\_Wigeon/South\_Carolina/22-000345-001/2021 (EPI\_ISL\_18133029) using MAFFT v7.526 \cite{Katoh2002}.
We generated two alignments by applying the following operations:
\begin{enumerate}[label=(\arabic*), topsep=1ex, itemsep=0.1ex, parsep=0.1ex]
  \item High-quality genomes (no ambiguous nucleotide sites) were selected.
  \item Data augmentation (ancestral sequence reconstruction) was applied.
  \item Duplicate sequences were removed, and among identical sequences, only the earliest (by collection date) was retained.
  \item Sequences in the alignment were arranged in time-reversed order (by collection date).
\end{enumerate}

\noindent\textbf{HA gene alignment:} \NumberOfHAGeneSequences distinct HA genes; length: \LengthOfHAGeneSequences nt; country: any; lineage: 2.3.4.4b; collection date: between December~2021 and November~2025.

\vspace{1mm}
\noindent\textbf{PB2 gene alignment:} \NumberOfPBtwoGeneSequences distinct PB2 genes; length: \LengthOfPBtwoGeneSequences nt; country: any; lineage: 2.3.4.4b; collection date: between December~2021 and November~2025.

\vspace{1mm}

\noindent For more details see Supplementary Information Section~1.2.

\subsubsection{Human immunodeficiency virus}

We used HIV-1 (strain HXB2) genome data provided by the European Nucleotide Archive (ENA) \cite{Leinonen2010} (for accession numbers see \hyperref[tab:supp_ena_accession_hiv_env]{Supplementary \autoref{tab:supp_ena_accession_hiv_env}}).
These data comprise \NumberOfHIVSequences envelope (Env) gene nucleotide sequences with genome coverage of at least 78\% (downloaded on 1~March~2025), which we aligned to the HIV1/HTLV-III/LAV reference genome (ENA accession number K03455.1) using MAFFT v7.526 \cite{Katoh2002}.
We generated an alignment by applying the following operations:
\begin{enumerate}[label=(\arabic*), topsep=1ex, itemsep=0.1ex, parsep=0.1ex]
  \item Genomes with no ambiguous nucleotide sites were selected.
  \item Data augmentation (ancestral sequence reconstruction) was applied.
  \item Duplicate sequences were removed.
\end{enumerate}

\noindent\textbf{Env gene alignment:} \NumberOfEnvGeneSequences distinct Env genes; length: \LengthOfEnvGeneSequences nt; country: any; lineage: HXB2; collection date: any.

\vspace{1mm}

\noindent For more details see Supplementary Information Section~1.3.

\subsection{Distance matrices}

We used the Python package Hammingdist v1.4.0 \cite{Keegan2024} to compute the \emph{genetic distance matrix} of each sequence alignment.
For any pair of sequences, this matrix encodes the \emph{Hamming distance} between the two sequences, which is the number of positions at which the nucleotides in the two aligned sequences differ.
Our convention in this work is that insertions and deletions (dashes ``-'' in aligned sequences) do not contribute to the genetic distance.
To gain efficiency, the genetic distance matrix was stored as a sparse matrix in coordinate format (COO).
In addition, since we were only interested in SNV cycles in persistent homology, it was sufficient to keep track of short distances only; all entries $> 3$ in the distance matrix were therefore reset to the value $3$.
Correspondingly, Hammingdist was called with the arguments \texttt{max\_distance=4} and \texttt{threshold=3}.
This reduced the time and memory usage required for storing genetic distances (\hyperref[fig:ripser]{Supplementary \autoref{fig:ripser}b-c}).
For the SARS-CoV-2 and H5N1 alignments, we applied the MuRiT algorithm \cite{Bleher2022} to incorporate temporal information into the distance matrix.
Every entry of~$1$ in the sparse distance matrix was replaced by the number of days between the date of the reference sequence and the most recent date in the pair of sequences corresponding to this entry in the distance matrix, and every entry of~$2$ ($3$) was replaced by the number of days between the date of the reference sequence and the most recent date occurring in the sequence alignment, plus $1$ (plus $2$).

\subsection{Persistent homology, persistence barcodes and topological cycles}

We used Ripser \cite{Ripser2021} to compute the persistent homology (persistence barcode and topological cycles) with $\mathbb{F}_2$ coefficients of each sequence alignment from its sparse genetic distance matrix.
Ripser is a state-of-the-art software for the computation of persistent homology based on the topological construction of Vietoris--Rips complexes (\hyperref[fig:ripser]{Supplementary \autoref{fig:ripser}a}).
For the computation of the persistence barcode, Ripser resorts to various optimizations, which are crucial when handling datasets of the size considered in this work.
Notably, Ripser computes persistent cohomology, which is not based on cycles but instead on cocycles, often described intuitively as \emph{cuts} that tear open a hole.
In order to obtain topological cycles representing the evolutionary features in persistent homology, we used a custom version of Ripser that subsequently carries out a second computation, this time based on cycles instead of cocycles.
This custom version of Ripser is available via commit f545200 on branch \texttt{tight-representative-cycles} of the GitHub repository \url{https://github.com/Ripser/ripser}.
While a naive computation based on homology would be prohibitively expensive, the previous computation of the persistence barcode based on cocycles makes the subsequent computation of representative cycles feasible.

The homological features (i.e.~bars in the persistence barcode) identified by persistent homology admit different cycle representatives in the dataset. In order to obtain topological cycles that fit tightly to the data points, the custom version of Ripser employs a method called \emph{exhaustive reduction} \cite{Edelsbrunner2019, Zomorodian2005}, which can be roughly summarized as follows:
Whenever a representative cycle contains an edge that also appears in another cycle as the longest edge, a tighter representative can be obtained by replacing the edge with the remaining edges from the other cycles, which all have shorter length.
The ambiguity of cycle representatives and bases has almost no effect on tRI, as we show in Supplementary Information Section~2.5.

We passed the sparse genetic distance matrix to Ripser with arguments \texttt{--dim 1 --format sparse --threshold 2}.
To enable time series analysis of sequence alignments, the fixed threshold of $2$ was replaced by the number of days between the date of the reference sequence and the most recent sequence date in the alignment, plus $1$.
Ripser returned a list of topological cycles, the edges of which are given by pairs of sequences.
Our choice of the \texttt{threshold} argument ensured that Ripser only returned those topological cycles in which all edges are of length~$1$, i.e.~SNV cycles.

We observed that Ripser performs faster if sequence alignments are sorted according to collection date in descending (i.e.~reverse chronological) order \cite{Bauer2022}.

\subsection{Topological recurrence analysis}

\emph{SNV cycles} are topological cycles all of whose edges correspond to SNVs (single nucleotide variations) in the sequence alignment.
Every edge in an SNV cycle has Hamming length $= 1$, and the endpoints of an edge correspond to a pair of sequences that differ at exactly one nucleotide site at which there is no dash ``-''.
Then for every SNV in the sequence alignment (defined with respect to the reference sequence), the \emph{raw topological recurrence index (raw tRI)} of this SNV is defined as the total number of SNV cycles in the sequence alignment that contain an edge corresponding to the SNV (\hyperref[fig:DefinitionOftRI]{\autoref{fig:DefinitionOftRI}c}).
The idea behind this definition is that the presence of a specific mutation in SNV cycles is indicative of convergent evolution in a twofold sense: (i) the mutation occurs independently twice in different lines of descent within each cycle, and furthermore, (ii) if the mutation is contained in different SNV cycles then it also occurs independently in different lines of descent within the phylogeny.

To compute the raw tRI of a given SNV, we analyzed the edges in the SNV cycles that had previously been found with Ripser.
Edges corresponding to the given SNV were counted under the conditions that (i) in each cycle, at most one such edge is counted, and (ii) no edge is counted twice.
This analysis was restricted to SNVs satisfying the following properties: (i) one of the two nucleotides involved in the SNV is the same as in the reference sequence at that site, and (ii) the other two nucleotides in the codon containing the SNV are the same as in the reference sequence.
These two conditions ensure that the corresponding SAAV is uniquely determined by the SNV with respect to the reference sequence.

In the case of the SARS-CoV-2 and H5N1 alignments which include collection dates, we obtained time series raw tRI values by recovering the temporal information from the edges of SNV cycles as follows.
The length of each edge in an SNV cycle encodes the most recent date of the two sequences defining that edge, given as the number of days counted from the date of the reference sequence.
This date was then assigned to the tRI signal contributed by the edge.

The final tRI was computed by applying two specific correction steps:
First, raw tRI values were adjusted by subtracting $1$ from every signal.
The motivation for this first correction step is the observation that a mutation present in precisely one SNV cycle in the dataset could have occurred just once in the evolutionary process (\hyperref[fig:DefinitionOftRI]{\autoref{fig:DefinitionOftRI}a}).
Only the presence of multiple SNV cycles containing a specific mutation is therefore indicative of the independent occurrence of that mutation.

Second, we normalized the adjusted raw tRI by its mean per nucleotide mutation type across four-fold degenerate sites, following the approach taken in \cite{Bloom2023a}.
The motivation for this second correction step is the nonuniformity of mutational spectra \cite{Bloom2023b}.
We first determined all nucleotide sites on the genome that are four-fold degenerate, i.e.~any of the four possible nucleotide variations at that site do not change the corresponding amino acid.
Then for each nucleotide mutation type, the mean adjusted raw tRI was computed across all four-fold degenerate SNVs in the sequence alignment (defined with respect to the reference sequence) that matched the given mutation type (\hyperref[fig:tri_fds]{Supplementary \autoref{fig:tri_fds}}).
To obtain the final tRI of an SNV, the raw tRI of that SNV was rescaled by dividing by this mean raw tRI associated with the mutation type of the given SNV.
If this mean raw tRI happened to be zero, the rescaling was not well-defined and no rescaling was applied.
In the case of time series raw tRI data, the rescaling was applied at every time step.
This may lead to spurious tRI signals at the onset of positive rescaling factors.

To focus the topological recurrence analysis on single genes or specific genomic regions (e.g.~the SARS-CoV-2 spike gene), we restricted the analysis to sliding windows on the genome.
For this, we first created a sub-alignment of genome sequences appropriately truncated to fit the sliding window, and then performed a separate topological recurrence analysis for this particular alignment.
This process typically leads to stronger tRI signals and more detailed tRI profiles (\hyperref[fig:tri_region_subalignment]{Supplementary \autoref{fig:tri_region_subalignment}a}).

We note that the time series topological recurrence analysis may exhibit an initial warm-up period, during which SNV cycles are formed with some delay.
This phenomenon can lead to spurious tRI growth rates and should be taken into account when interpreting early signals in tRI activity landscapes.

\subsection{Downstream analysis}

We analyzed time series tRI signals for the SARS-CoV-2 whole genome and spike gene alignments (\hyperref[tab:supp_tri_whole_genome]{Supplementary Table \ref{tab:supp_tri_whole_genome}}), both for individual mutations and across genes on the genome.
For individual SNVs, we generated \emph{tRI/prevalence time series charts} showing time-dependent tRI, together with the conservative tRI significance level, and time-dependent prevalence of the SNV (\hyperref[fig:DynamicsOftRI]{\autoref{fig:DynamicsOftRI}a}).
In these charts, the horizontal axis represents the time range at daily resolution.
The time-dependent prevalence was computed as the quotient of the number of sequences carrying the SNV by the total number of sequences in the alignment, up to a given point in time.

We generated a \emph{dynamic tRI landscape} from SARS-CoV-2 spike gene time series tRI data to comprehensively map time-dependent tRI signals across this gene (\hyperref[fig:DynamicsOftRI]{\autoref{fig:DynamicsOftRI}b}).
This tRI landscape is a two-dimensional heat map, where the horizontal axis represents the amino acid positions on the gene (from left to right), the vertical axis represents the time range (from top to bottom) and the heat parameter is the log-transformed tRI.
Vertical lines in this heat map represent time-dependent tRI signals at a specific amino acid site.
For this, we aggregated tRI signals of non-synonymous SNVs into tRI signals of amino acid sites by summing up tRI signals of all SNVs that give rise to SAAVs at the same amino acid site.
We further derived a \emph{dynamic tRI activity landscape} from this tRI landscape, which is a two-dimensional heat map where the heat parameter is the positive part of the tRI growth rate (the first discrete differences of the exponentially weighted moving average of the log-transformed tRI) in the vertical direction.
We generated dynamic tRI activity landscapes of the SARS-CoV-2 spike gene (\hyperref[fig:DynamicsOftRI]{\autoref{fig:DynamicsOftRI}c}) and UK spike gene (\hyperref[fig:Adaptation]{\autoref{fig:Adaptation}c}) alignments and for the H5N1 PB2 gene alignment (\hyperref[fig:Adaptation]{\autoref{fig:Adaptation}g}).
For better visualization and to enhance hotspots of convergent evolution, we applied two-dimensional Gaussian smoothing with standard deviation~$0.4$~($3$) in the horizontal (vertical) direction in the spike gene analysis, and with standard deviation~$4$~($10$) in the horizontal (vertical) direction in the UK spike gene and the PB2 gene analyses. 
For the SARS-CoV-2 spike gene, we took horizontal slices in dynamic tRI (activity) landscapes to extract \emph{snapshots} showing landscapes of tRI (tRI growth rates) across the spike gene at specific points in time.
For better visualization we applied one-dimensional Gaussian smoothing with standard deviation $2$.

\subsection{Statistical analyses}

Topological cycles can arise randomly in neutral evolution, causing some background noise in tRI (\hyperref[fig:tri_simulated_scenarios_stats]{\autoref{fig:tri_simulated_scenarios_stats}a} and \hyperref[fig:topological_noise]{Supplementary \autoref{fig:topological_noise}}).
To quantitatively estimate the effect of this background noise in SARS-CoV-2 whole genome tRI signals, we simulated neutral evolutionary scenarios using a Wright-Fisher forward model of viral evolution using SANTA-SIM v1.0 \cite{Jariani2019} with (i) fixed parameters: genome length ($30,000$ nt), number of generations (${N = 10,000}$), number of sequences sampled from the population per time step (${n=15}$), recombination rate (${\rho = 0}$), transition bias (1.0), and (ii) variable parameters: mutation rate per site per generation~${\mu}$, population size~$p$, carrying population~$c$, population growth rate per generation~$g$.
We considered five scenarios: In scenarios I-III we varied the mutation rate $\mu$ under the assumption of fixed population size $p$, while in scenarios IV and V we investigated the effects of logistic growth $g$ of the population with carrying population $c$.
For more details see Supplementary Information Section~2.1.

We analyzed the effect of stochastic sequencing errors on tRI by simulating stochastic sequencing errors with varying error rates in two reference alignments (\hyperref[fig:tri_simulated_scenarios_stats]{\autoref{fig:tri_simulated_scenarios_stats}b} and \hyperref[fig:sequencing_errors]{Supplementary \autoref{fig:sequencing_errors}}).
As reference alignments, we used a synthetic alignment simulated under neutral evolution with SANTA-SIM v1.0 \cite{Jariani2019}, and a specially tailored real SARS-CoV-2 200k whole genomes alignment.
We swept sequencing error rates from $10^{-8}$ to $10^{-3}$ on a logarithmic scale.
For more details see Supplementary Information Section~2.2.

We applied a Kolmogorov–Smirnov test to compare the frequency distributions of raw tRI signals in the SARS-CoV-2 whole genome and the spike gene alignments with the frequency distributions obtained from a Monte Carlo permutation test under the null hypothesis that raw tRI signals are distributed uniformly across the genome.
For more details see Supplementary Information Section~2.3.

These findings led us to introduce a conservative tRI significance level, by means of a permutation test under the null hypothesis of uniformly distributed raw tRI signals, with $p = 0.05$ as the threshold for statistically significant signals.
For more details see Supplementary Information Section~2.4.

\subsection{Validation on simulated data}

We analyzed the sensitivity of the tRI for positive selection by varying fitness and sampling density in synthetic nucleotide sequence alignments generated with SANTA-SIM v1.0 \cite{Jariani2019}.
For~$1.7\%$ of single nucleotide mutations, we varied the fitness parameter between~$0.98$ and~$1.10$ in steps of $0.02$ (variants with varied fitness), while all other mutations were assigned a fitness parameter of $1.00$ (neutral variants).
Fitness values in this range are typical for single nucleotide substitutions in RNA viruses \cite{Sanjun2004}.
For each scenario, we computed the mean tRI per SNV separately across all variants with varied fitness and across all neutral variants.
We then determined the Spearman correlation between fitness and mean tRI signals across variants with varied fitness.
Next, we kept the fitness of variants with varied fitness fixed at~$1.10$ (beneficial variants) and randomly selected $10\%, 20\%, \ldots, 100\%$ of sequences in simulated alignments.
For each scenario, we computed the effect size (Cohen's $d$) for the difference between mean tRI across beneficial variants and mean tRI across neutral variants.
For more details see Supplementary Information Section~3.1.

We compared raw tRI signals with phylogenetic homoplasy counts using synthetic nucleotide sequence alignments generated with SANTA-SIM~v1.0 \cite{Jariani2019}.
For~$7.1\%$ of single nucleotide mutations, we increased the fitness parameter to $1.3$ (beneficial variants), while all other mutations were assigned a neutral fitness parameter of $1.0$ (neutral variants).
For each simulated alignment, we (i) removed duplicate sequences and computed raw tRI signals with EVOtRec, and (ii) reconstructed phylogenetic trees with IQ-TREE v3.0.1 \cite{Wong2025} and subsequently computed homoplasy counts with TreeTime v0.11.4~\cite{Sagulenko2018}.
We then determined the Spearman correlation between raw tRI profiles and phylogenetic homoplasy counts, separately for profiles across neutral variants and for profiles across beneficial variants.
For more details see Supplementary Information Section~3.2.

\subsection{Validation on experimental data}

We validated tRI signals of amino acid changes in the SARS-CoV-2 (Omicron) spike gene alignments (\hyperref[tab:supp_tri_whole_genome]{Supplementary Table \ref{tab:supp_tri_whole_genome}}) and in the H5N1 HA and PB2 gene alignments (\hyperref[tab:supp_tri_influenza_ha]{Supplementary Table \ref{tab:supp_tri_influenza_ha}}) on published deep mutational scanning (DMS) data.
For this, we aggregated tRI signals of non-synonymous single nucleotide variants (SNV) into tRI signals of single amino acid variants (SAAV) by summing up the tRI of all SNVs that give rise to the same SAAV.
We computed the time-dependent Pearson correlation between SARS-CoV-2 spike gene tRI and DMS effect \cite{Starr2022a}.
For more details see Supplementary Information Section~4.1.

Moreover, we compared tRI signals in the SARS-CoV-2 Omicron spike gene alignment with DMS effects for the BA.1 \cite{Dadonaite2023}, BA.2 \cite{Dadonaite2024a} and XBB.1.5 \cite{Dadonaite2024a} sub-lineages (\hyperref[fig:Adaptation]{\autoref{fig:Adaptation}b} and \hyperref[fig:stats_tri_fitness_dms]{Supplementary \autoref{fig:stats_tri_fitness_dms}a-d}), based on $2 \times 2$ contingency tables for binarized tRI ($\textrm{tRI} > 0$ vs.~$\textrm{tRI} = 0$) and binarized DMS effect ($\textrm{DMS effect} > 0$ vs.~$\textrm{DMS effect} \le 0$).
In addition, for XBB.1.5 we computed a logistic regression model predicting positive tRI as a function of the three XBB.1.5 DMS effects.
We repeated all analyses with tRI replaced by published phylogeny-based fitness effect estimates from \cite{Bloom2023a, Haddox2025}.
In this case, for XBB.1.5 we computed an ordinary least squares regression model predicting positive fitness effect as a function of the three XBB.1.5 DMS effects.
For more details see Supplementary Information Section~4.2.

Similarly, we compared tRI signals obtained from the H5N1 HA and PB2 gene alignments with DMS effects for the HA gene \cite{Dadonaite2024b} and PB2 gene \cite{Soh2019} (\hyperref[fig:Adaptation]{\autoref{fig:Adaptation}d}), based on $2 \times 2$ contingency tables for binarized tRI ($\textrm{tRI} > 0$ vs.~$\textrm{tRI} = 0$) and binarized DMS effect ($\textrm{DMS effect} > 0$ vs.~\mbox{$\textrm{DMS effect} \le 0$}).
For more details see Supplementary Information Section~4.3.

\subsection{Comparison with phylogeny-based fitness effect estimates}

We compared EVOtRec tRI signals with published phylogeny-based fitness effect estimates \cite{Bloom2023a} for SNVs on the SARS-CoV-2 whole genome and spike gene (\hyperref[fig:Benchmarking]{\autoref{fig:Benchmarking}c-d}).
We log-transformed positive tRI scores of the SARS-CoV-2 whole genome and spike gene alignments (\hyperref[tab:supp_tri_whole_genome]{Supplementary Table \ref{tab:supp_tri_whole_genome}}).
Fitness effect estimates were taken from the column ``delta\_fitness'' in \cite{BloomNeherData2024a}.
We subsequently computed the Pearson correlation between log-transformed positive tRI and fitness effect estimates, separately for non-synonymous and synonymous SNVs on the whole genome (resp.~spike gene) for which both tRI and fitness effect estimate data were available.

\subsection{Performance analysis and benchmarking}

We analyzed the runtime and memory usage of EVOtRec in comparison with state-of-the-art tools for the reconstruction of phylogenetic trees (\hyperref[fig:Benchmarking]{\autoref{fig:Benchmarking}a-b}).
We applied the following tools to eight SARS-CoV-2 whole genome alignments with up to 200k distinct sequences, so long as the runtime did not exceed 24 hours:
The Python script evotrec.py v1.0, available via \url{https://github.com/ottamj/evotrec}; IQ-TREE v3.0.1 \cite{Wong2025}; VeryFastTree v4.0.5 \cite{Pieiro2024}; CMAPLE v1.1.0 \cite{LyTrong2024} and UShER v0.6.6 \cite{Turakhia2021}.
To obtain the sequence alignments, we sorted the SARS-CoV-2 whole genome alignment according to collection date in ascending order and selected the first $n$ sequences, for $n = 1\times10^3, 2\times10^3, 5\times10^3, 1\times10^4, 2\times10^4, 5\times10^4, 1\times10^5, 2\times10^5$.
We analyzed the runtime and peak memory usage (RAM), computing mean ± 95\% confidence interval (CI) over \NumberRunsBenchmarking independent runs for each tool and each alignment.
All computations were performed on a server with Intel Xeon Gold 6230R processors (2.10GHz) and 52 cores.
For more details see Supplementary Information Section~5.

\subsection{Data availability}
The findings of this study are based (i) on genome data of \NumberOfGISAIDCoronaSequences SARS-CoV-2 nucleotide sequences available from the GISAID EpiCoV Database as of 24~February~2024, via \url{https://doi.org/10.55876/gis8.240314pc}, (ii) on genome data of \NumberOfGISAIDInfluenzaHASequences H5N1 HA gene nucleotide sequences available from the GISAID EpiFlu Database as of 16~November~2025, via \url{https://doi.org/10.55876/gis8.251116eh}, (iii) on genome data of \NumberOfGISAIDInfluenzaPBtwoSequences H5N1 PB2 gene nucleotide sequences available from the GISAID EpiFlu Database as of 16~November~2025, via \url{https://doi.org/10.55876/gis8.251116bc}, and (iv) on genome data of \NumberOfHIVSequences HIV-1 Env gene nucleotide sequences available from the European Nucleotide Archive (ENA) as of 1~March~2025, via \url{https://www.ebi.ac.uk/ena/browser/home} using the accession numbers provided in \hyperref[tab:supp_ena_accession_hiv_env]{Supplementary \autoref{tab:supp_ena_accession_hiv_env}}.
Experimental data (deep mutational scanning) of the SARS-CoV-2 wild-type RBD were taken from the published study \cite{Starr2022a} and are available from \cite{StarrDMS2022}.
Experimental data (deep mutational scanning) of the SARS-CoV-2 Omicron spike gene were taken from the published studies \cite{Dadonaite2023, Dadonaite2024a} and are available from \cite{DadonaiteDMS2023} for BA.1, from \cite{DadonaiteDMS2024ba2} for BA.2 and from \cite{DadonaiteDMS2024xbb} for XBB.1.5.
Data on SARS-CoV-2 fitness effect estimates were taken from the published studies \cite{Bloom2023a, Haddox2025} and are available from \cite{BloomNeherData2024a, BloomNeherData2024b} and \cite{HaddoxData2025_21K, HaddoxData2025_BA2, HaddoxData2025_XBB}.
Experimental data (deep mutational scanning) of the avian influenza A subtype H5N1 HA and PB2 genes were taken from the published studies \cite{Dadonaite2024b, Soh2019} and are available from \cite{DadonaiteDMS2024h5n1, SohDMS2019}.
Results of the topological recurrence analysis (tRI and prevalence data) for SARS-CoV-2, H5N1 and HIV-1 are available via \url{https://doi.org/10.5281/zenodo.18332026}.

\subsection{Code availability}
Code of the EVOtRec pipeline is available via \url{https://github.com/ottamj/evotrec}.
Code of Hammingdist and Ripser is available via \url{https://github.com/ssciwr/hammingdist} and \url{https://github.com/Ripser/ripser/tree/tight-representative-cycles}.
All other code is available from the corresponding authors upon request.

\section{Acknowledgements}
We gratefully acknowledge all data contributors, i.e., the Authors and their Originating laboratories responsible for obtaining the specimens, and their Submitting laboratories for generating the genetic sequence and metadata and sharing via the GISAID Initiative and the European Nucleotide Archive (ENA), on which this research is based.
This work was supported by the de.NBI Cloud within the German Network for Bioinformatics Infrastructure (de.NBI) and ELIXIR-DE (Forschungszentrum Jülich and W-de.NBI-001, W-de.NBI-004, W-de.NBI-008, W-de.NBI-010, W-de.NBI-013, W-de.NBI-014, W-de.NBI-016, W-de.NBI-022).
The authors further acknowledge support from the Interdisciplinary Center for Scientific Computing at Heidelberg University and the Scientific Software Center of Heidelberg University.
This research was supported by the DFG Collaborative Research Center SFB/TRR 109 ``Discretization in Geometry and Dynamics''.
M.B.~was supported by the Deutsche Forschungsgemeinschaft (DFG, German Research Foundation) under Germany's Excellence Strategy EXC 2181/1 - 390900948 (the Heidelberg STRUCTURES Excellence Cluster).
L.H.~thanks the Evangelisches Studienwerk Villigst for their support.
A.O.~and M.N.~were supported by the Vector Stiftung.
M.N.~was supported by the Helmholtz Association under the joint research school ``HIDSS4Health – Helmholtz Information and Data Science School for Health''.
A.O.~acknowledges funding by the Deutsche Forschungsgemeinschaft (DFG, German Research Foundation) -- 281869850 (RTG 2229).
This research was funded by the Federal Ministry of Education and Research (BMBF) and the Baden-W\"urttemberg Ministry of Science as part of the Excellence Strategy of the German Federal and State Governments.
Work conducted by Z.A.~at the Wellcome Sanger Institute was supported by core funding from the Wellcome Trust (Grant No.~220540/Z/20/A).

\section{Author contributions}
M.B., L.H., M.N., Z.A., J.P.-G., R.R., A.O.~designed the study;
A.O. managed and supervised the project;
M.B., L.H., A.O.~curated data;
M.B., L.H., M.N, Z.A., M.C., J.P.-G., A.O.~performed computational analyses;
M.B., L.H., M.N., U.B., A.O.~developed and implemented software;
M.B., L.H., A.O.~acquired computing resources;
M.B., L.H., M.N., Z.A., U.B., R.R., A.O.~wrote the manuscript;
M.B., L.H., M.N., A.O.~designed and implemented the EVOtRec pipeline;
M.N.~designed the MuRiT algorithm for time series topological recurrence analysis;
A.O.~designed and implemented tRI/prevalence charts and dynamic tRI landscapes;
M.B., M.N., A.O.~implemented the MuRiT algorithm in the EVOtRec pipeline;
U.B.~designed and implemented exhaustive reduction for cycle localization in Ripser;
M.B~designed and implemented analyses of simulated evolutionary scenarios;
M.B., A.O.~performed simulations;
A.O.~designed, implemented and performed benchmarking and comparison with phylogenetic methods;
M.B, A.O.~performed analyses of simulated evolutionary scenarios;
M.N, A.O.~designed stochastic sequencing error analysis;
A.O.~performed stochastic sequencing error analysis;
M.N.~designed and implemented ancestral sequence reconstruction for alignments with low sampling rate;
M.N.~designed and implemented study of effect of ambiguity of cycle representatives;
A.O.~analyzed topological signals at four-fold degenerate sites;
A.O.~performed time series topological recurrence analyses of genomic datasets;
M.N., A.O.~designed comparison with experimental data;
A.O.~performed comparison with experimental data;
M.B., L.H., M.N., Z.A., U.B., A.O.~designed and created figures; all authors contributed to the final version of the paper.

\section{Competing interests}
R.R.~is a founder of Genotwin, he is member of the Scientific Advisory Board of AimedBio and consults for Arquimea Research.
The other authors declare no competing interests.

\section{Supplementary Material}

\begin{itemize}
  \item Supplementary Tables
  \item Supplementary Figures
  \item Supplementary Information
\end{itemize}

\noindent \textbf{\sffamily{Correspondence and requests for materials}} should be addressed to M.B., M.N., R.R. or A.O.

\printbibliography[resetnumbers=true]

\clearpage
\onecolumn

\setcounter{figure}{0}
\setcounter{table}{0}
\makeatletter
\renewcommand{\fnum@figure}{Supplementary Figure \thefigure}
\renewcommand{\fnum@table}{Supplementary Table \thetable}
\let\c@figure\c@table
\makeatother

\captionsetup[table]{singlelinecheck=false}

\section{Supplementary Material}

\label{sec:supp_info}

\begin{table}[H]
  \centering
	\includegraphics[width=15cm]{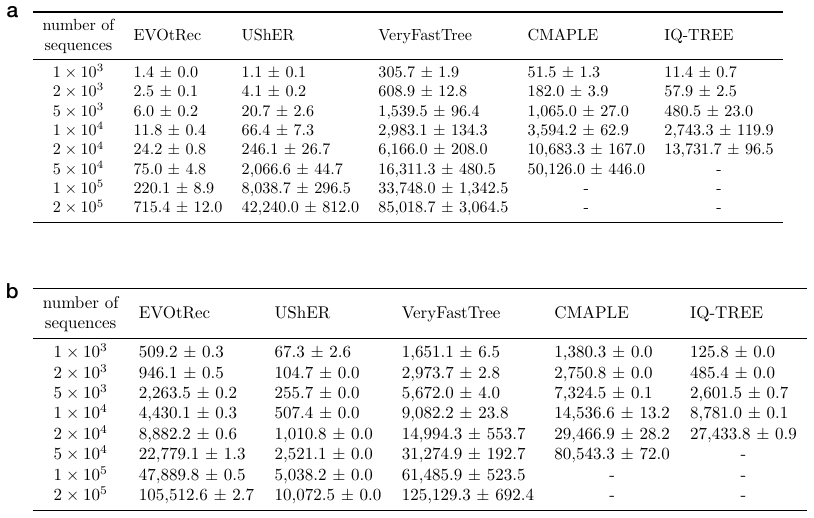}
	\caption{\textbf{\sffamily{Benchmarking of EVOtRec vs.~phylogenetic inference tools.}}
We analyzed the runtime and memory usage of EVOtRec in comparison with the following tools for the reconstruction of phylogenetic trees: UShER \cite{Turakhia2021}, VeryFastTree \cite{Pieiro2024}, CMAPLE \cite{LyTrong2024} and IQ-TREE \cite{Wong2025}.
We applied these tools to eight SARS-CoV-2 genome alignments with 1k up to 200k distinct sequences, obtained by subsampling the SARS-CoV-2 whole genome alignment (\nameref{methods}).
Tables show the mean ± 95\% CI (over~\NumberRunsBenchmarking independent runs) of runtime in seconds (a) and peak memory usage (RAM) in megabytes (b).
A dash (-) indicates that the runtime exceeded 24 hours, in which case no measurement was taken.
}
\label{tab:tri_table_benchmarking}
\end{table}

\begin{table}[H]
\captionabove{\textbf{\sffamily{Topological recurrence analysis of SARS-CoV-2.}}
Time series data (raw tRI, tRI, tRI significance level, prevalence, GISAID accession numbers) of the topological recurrence analysis of the following SARS-CoV-2 alignments: (i) whole genome alignment (starting date 30 December 2019); (ii) spike gene alignment (starting date 30 December 2019); (iii) Omicron spike gene alignment (starting date 01 November 2021); (iv) United Kingdom spike gene alignment (starting date 30 December 2019).
CSV data files are available via \url{https://doi.org/10.5281/zenodo.18332026}.
}
\label{tab:supp_tri_whole_genome}
\label{tab:supp_tri_spike_gene}
\label{tab:supp_tri_omicron_spike_gene}
\label{tab:supp_tri_uk_spike_gene}
\end{table}

\begin{table}[H]
\captionabove{\textbf{\sffamily{Topological recurrence analysis of H5N1.}}
Time series data (raw tRI, tRI, tRI significance level, prevalence, GISAID accession numbers) of the topological recurrence analysis of the following H5N1 alignments: (i) HA gene alignment (starting date 30 December 2021); (ii) PB2 gene alignment (starting date 30 December 2021).
CSV data files are available via \url{https://doi.org/10.5281/zenodo.18332026}.
}
\label{tab:supp_tri_influenza_ha}
\label{tab:supp_tri_influenza_pb2}
\end{table}

\begin{table}[H]
\captionabove{\textbf{\sffamily{Topological recurrence analysis of HIV-1.}}
Data (raw tRI, tRI, tRI significance level, prevalence, ENA accession numbers) of the topological recurrence analysis of the HIV-1 Env gene alignment.
CSV data files are available via \url{https://doi.org/10.5281/zenodo.18332026}.
}
\label{tab:supp_tri_hiv_env}
\label{tab:supp_ena_accession_hiv_env}
\end{table}

\begin{table}[H]
  \centering
	\includegraphics[width=16cm]{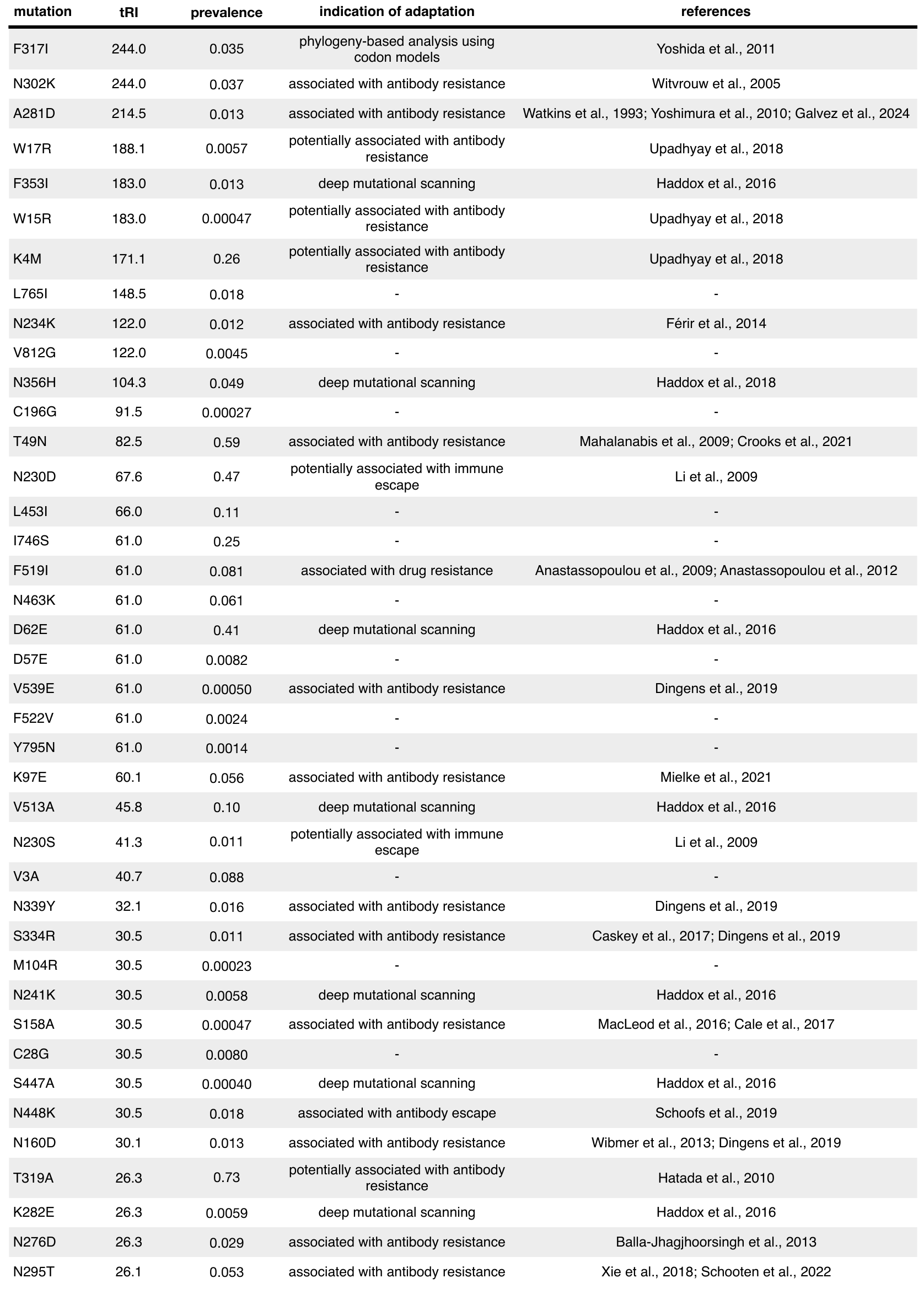}
	\caption{\textbf{\sffamily{Assessment of adaptation for HIV-1 amino acid changes with strong topological signal.}}
The table displays the top~40 amino acid changes on the HIV-1 envelope gene (Env) with strongest tRI signals in the EVOtRec analysis of the ENA Env gene alignment \cite{Leinonen2010}, together with their prevalence in the sequence alignment (\hyperref[tab:supp_tri_hiv_env]{Supplementary Table~\ref{tab:supp_tri_hiv_env}} and \nameref{methods}).
Among these, at least 70\% (28 out of 40) have been associated with (potential) adaptation in the literature \cite{Yoshida2011, Witvrouw2005, Watkins1993, Yoshimura2010, Galvez2024, Upadhyay2018, Haddox2016, Ferir2014, Mahalanabis2009, Crooks2021, Li2009, Anastassopoulou2009, Anastassopoulou2012, Mielke2021, Caskey2017, MacLeod2016, Cale2017, Schoofs2019, Wibmer2013, BallaJhagjhoorsingh2013, Xie2018, vanSchooten2022, Hatada2010, Haddox2018, Dingens2019}.
}
\label{tab:tri_table_hiv}
\end{table}

\setcounter{figure}{0}

\begin{figure}[h!]
  \centering
	\includegraphics[width=\textwidth]{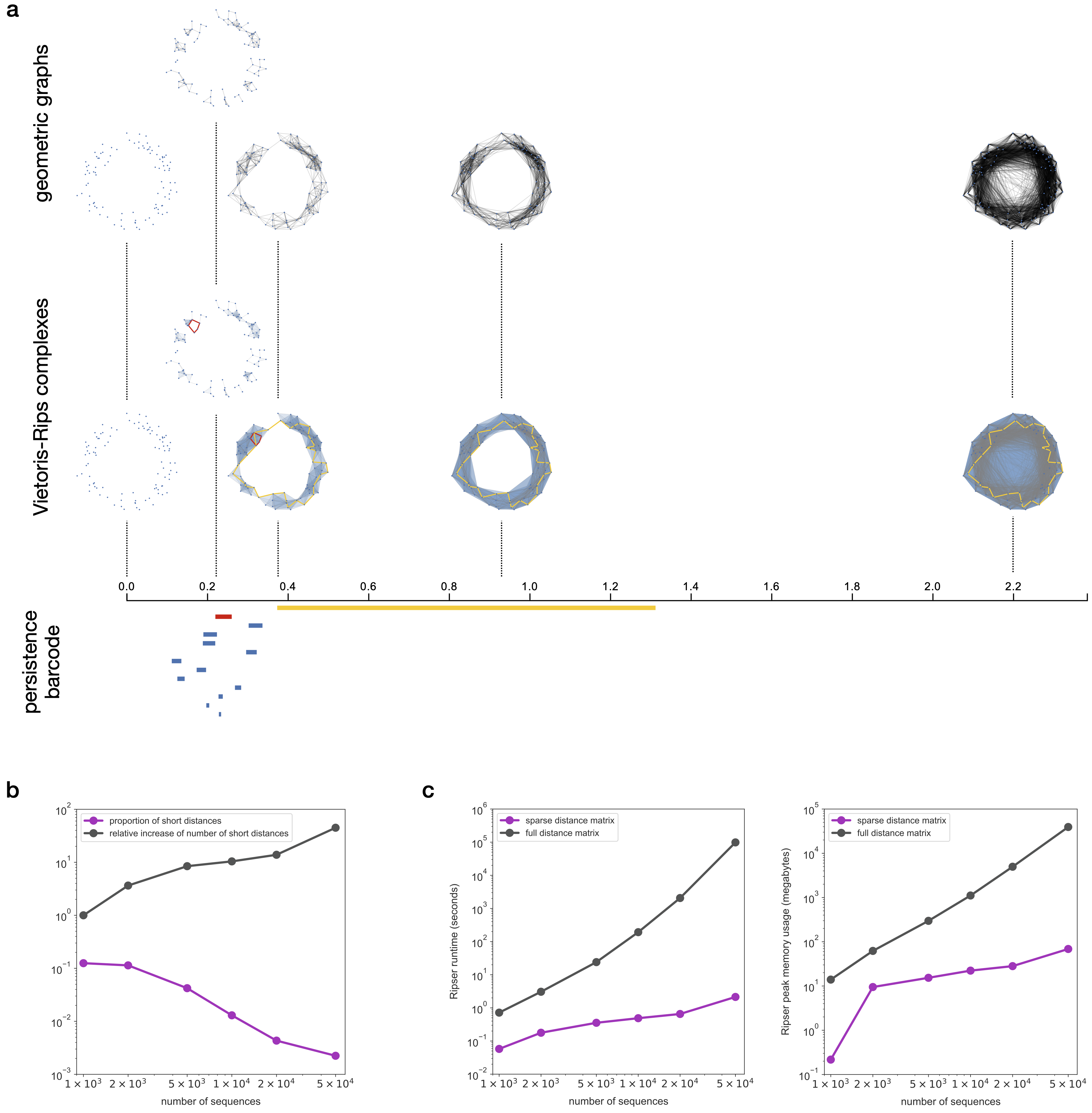}
	\caption{\textbf{\sffamily{Computing persistent homology of genomic datasets.}}
EVOtRec uses Ripser \cite{Bauer2021} to compute the persistent homology of genomic datasets from genetic distance matrices (\nameref{methods}).
\textbf{(a)} Schematic example of how Ripser can be used to study the topology of genomic datasets.
Each point represents a sample, and we display the geometric graphs (upper row), the resulting Vietoris--Rips complexes at different scales (middle row) and the persistence barcode in dimension one (lower row).
If one only chooses one scale, one might either see nothing, or detect the small red cycle but miss the large yellow one, or vice versa.
The basic idea of persistent homology to handle this issue is to characterize each cycle with its scale of appearance and disappearance: the red cycle induces a red bar in the barcode, and similarly for the yellow cycle.
\textbf{(b, c)} EVOtRec's sparse distance format reduces the complexity of persistent homology computations.
The format is based exclusively on short genetic distances, whose proportion among all observed distances rapidly decreases to less than 1\% for alignments with more than 10k sequences~(b).
The usage of EVOtRec sparse distance matrices significantly reduces runtime and peak memory usage (RAM) for the computation of persistence barcodes with Ripser by several orders of magnitude as compared with standard full distance matrices (c).
Runtimes in (c) are means over 5 independent runs.
}
\label{fig:ripser}
\end{figure}

\begin{figure}[h!]
  \centering
	\includegraphics[width=\textwidth]{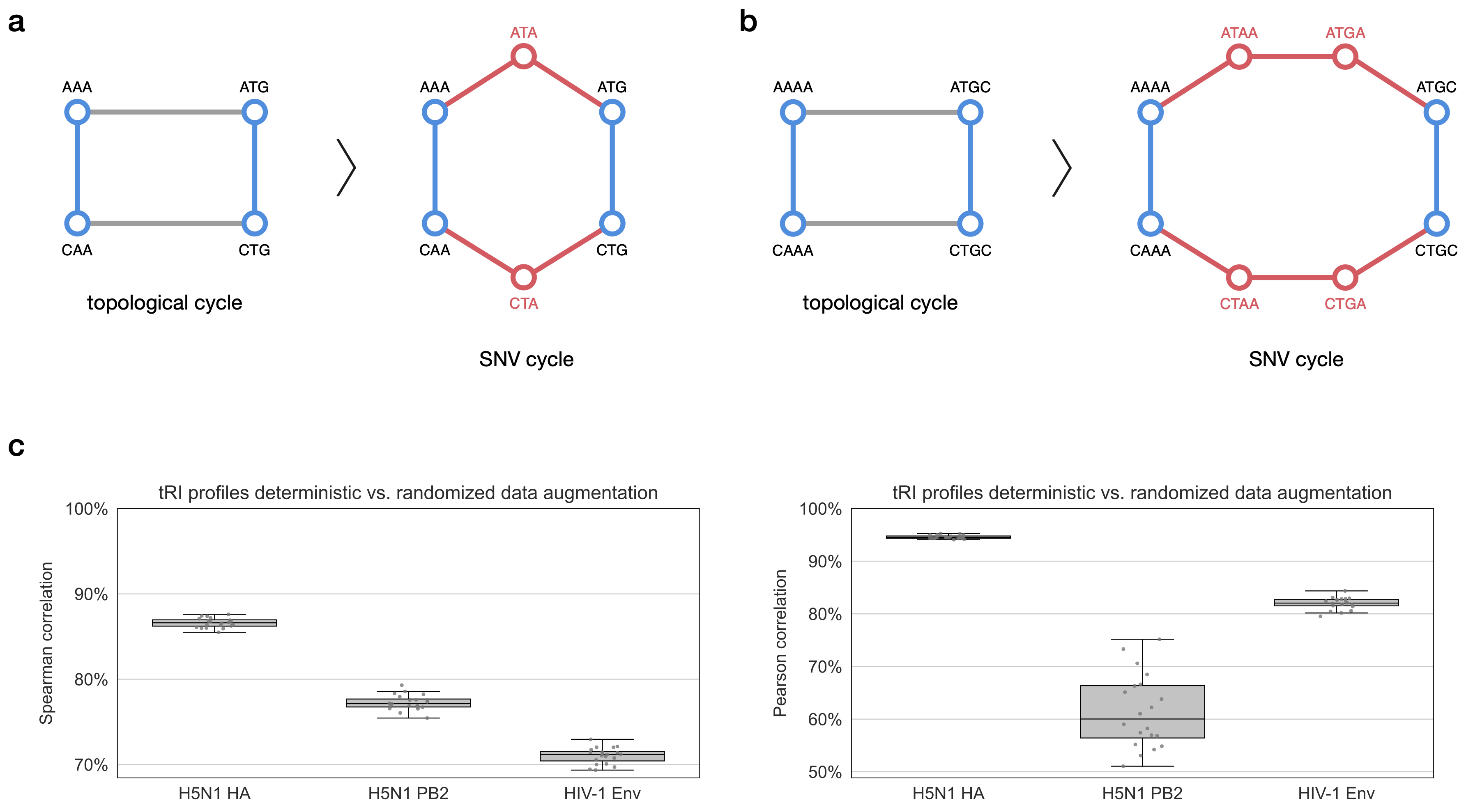}
	\caption{\textbf{\sffamily{Data augmentation via ancestral sequence reconstruction.}}
We applied data augmentation via ancestral sequence reconstruction in the H5N1 and HIV-1 genome alignments, as these alignments have lower sampling rate than the SARS-CoV-2 alignments (\nameref{methods}).
\textbf{(a-b)} 
For short edges (Hamming length = 2 or 3), synthetic sequences were added to the alignment to replace these edges with chains composed exclusively of unit-length edges.
The procedure turns topological cycles containing edges of length~2 or~3 into SNV cycles, thereby amplifying the corresponding tRI signals.
In the displayed schematic example with trinucleotide sequences (a), by adding two synthetic sequences (red) each of the horizontal edges of Hamming length~2 in the topological cycle is replaced by two unit-length edges (red), turning the topological cycle into an SNV cycle.
In the displayed schematic example with tetranucleotide sequences (b), by adding four synthetic sequences (red) each of the horizontal edges of Hamming length~3 in the topological cycle is replaced by a chain of three unit-length edges (red), turning the topological cycle into an SNV cycle.
\textbf{(c)} Box plot with overlaid strip plot showing the Spearman and Pearson correlation between tRI profiles generated with deterministic (as described in a, b) versus randomized data augmentation (\nameref{methods}), for the H5N1 and HIV-1 alignments.
Correlations were computed between a single alignment that was generated using deterministic data augmentation, and 20 alignments that were independently generated using randomized data augmentation.
We found strong correlations across all three alignments, justifying the use of data augmentation in the preprocessing of the H5N1 and HIV-1 alignments.
}
\label{fig:data_augmentation}
\end{figure}

\begin{figure}[h!]
  \centering
  \includegraphics[width=\textwidth]{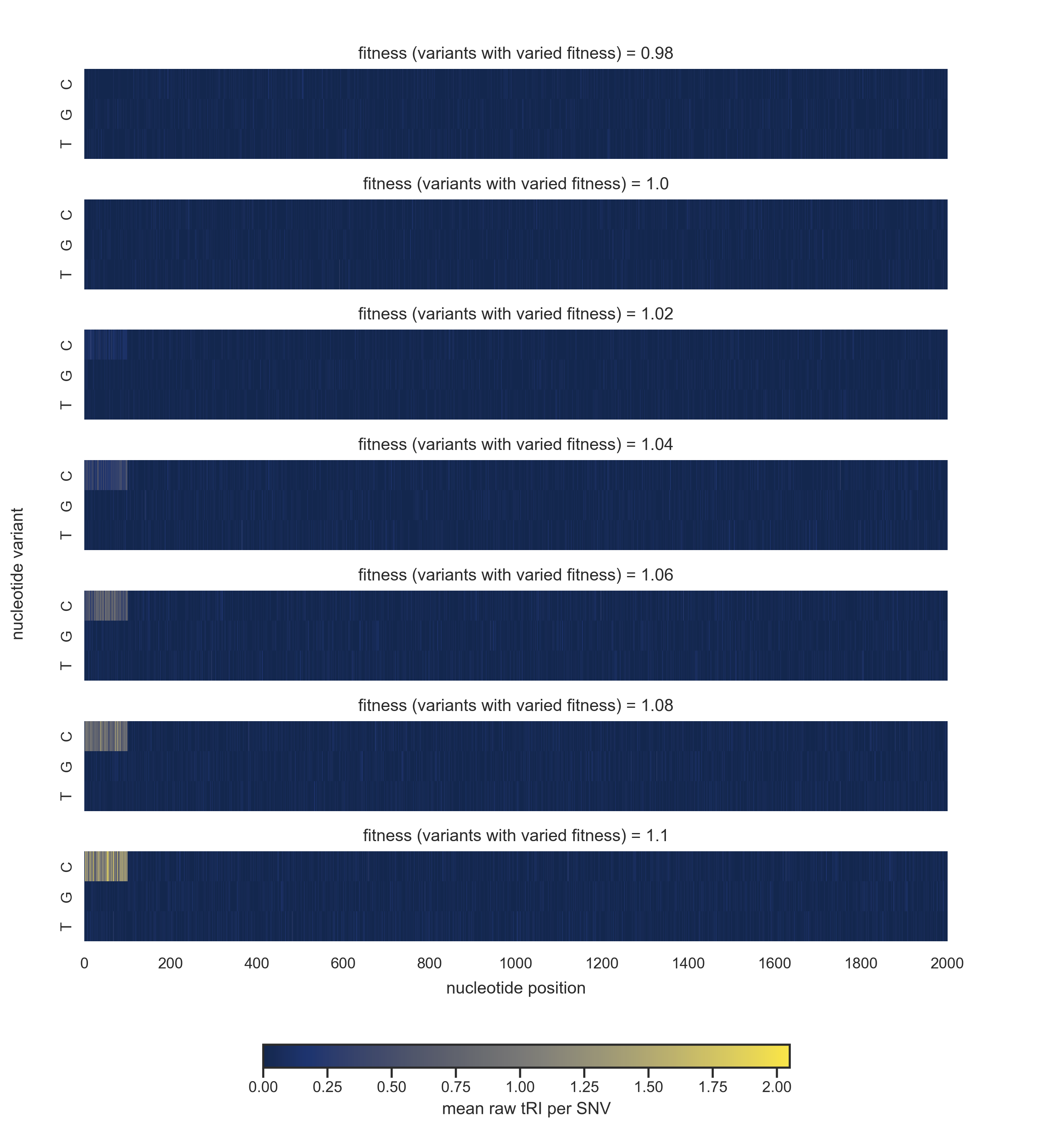}
  \caption{
\textbf{\sffamily{Effects of variations in fitness on tRI signals.}}
We analyzed the dependence of the raw tRI on the fitness parameter in simulated synthetic sequence alignments generated with SANTA-SIM v1.0 \cite{Jariani2019}.
Simulations were initialized with polyadenine sequences AA\ldots A of length 2000nt (\nameref{methods}).
The fitness parameter assigned to~C alleles at the first $100$ nucleotide positions (variants with varied fitness) was uniformly set to one of the values $0.98$, $1.00$, $1.02$, $1.04$, $1.06$, $1.08$ and $1.10$, making the mutations A1C, A2C, \ldots, A100C deleterious, neutral or beneficial, respectively.
The fitness parameter for all other mutations was kept at $1.00$ (neutral variants). For each of these scenarios we ran 20 independent simulations.
We performed a topological recurrence analysis with EVOtRec for each simulated alignment and subsequently computed the mean raw tRI per SNV (mean calculated by averaging over all 20 simulations for each SNV).
The displayed heatmaps show the mean raw tRI per SNV, with the horizontal axis representing the~2000 nucleotide positions and the vertical axis representing the three mutation types A\textgreater C, A\textgreater G and A\textgreater T.
}
\label{fig:tri_simulated_scenarios}
\end{figure}

\begin{figure}[h!]
  \centering
	\includegraphics[width=\textwidth]{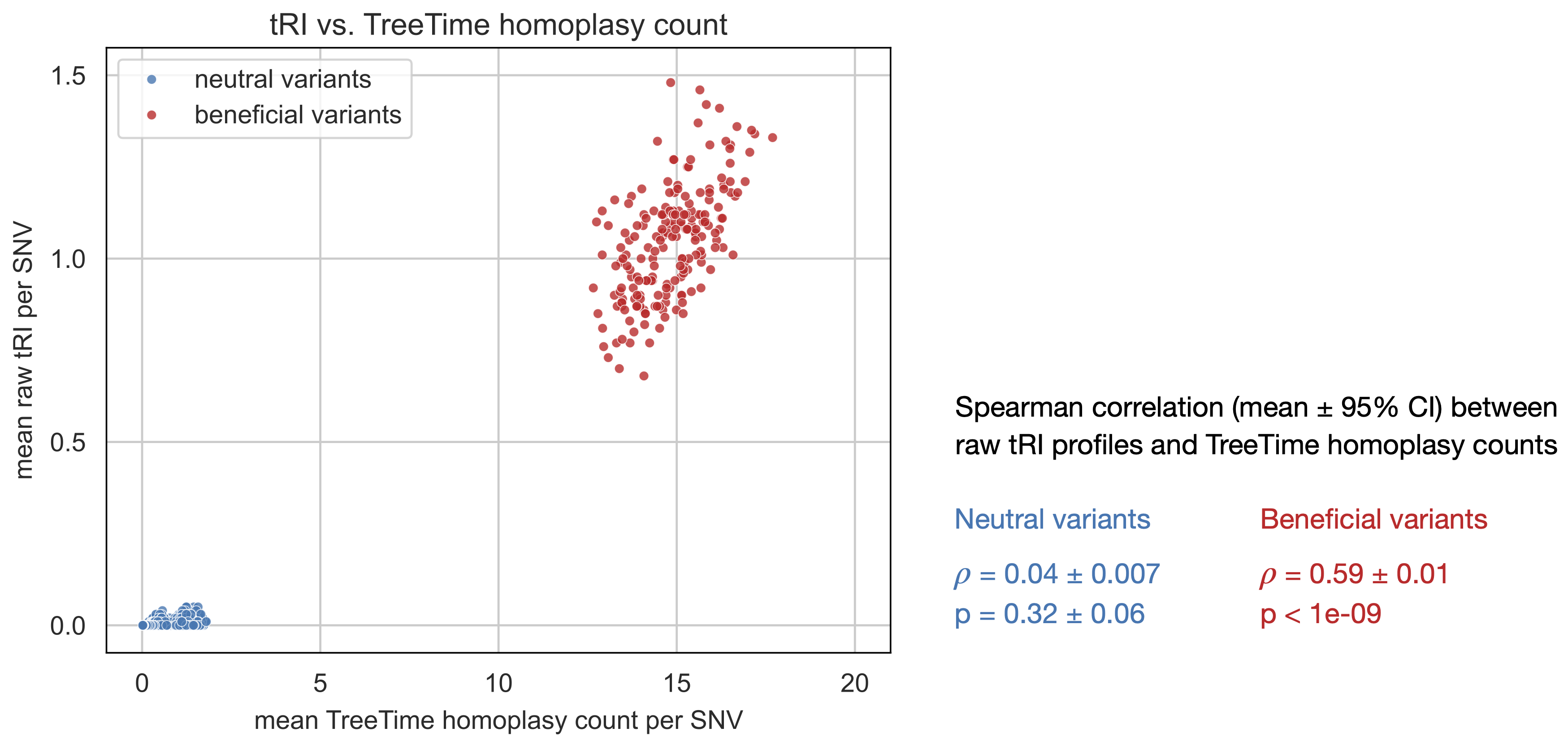}
	\caption{\textbf{\sffamily{Comparison of tRI and phylogenetic homoplasy counts in simulated evolutionary scenarios.}}
We simulated~100 nucleotide sequence alignments with SANTA-SIM~v1.0 \cite{Jariani2019}, where $7.1\%$ of single nucleotide mutations were assigned a fitness parameter of $1.3$ (beneficial variants), while all other mutations were assigned a fitness parameter of $1.0$ (neutral variants) (\nameref{methods}).
For each alignment, we (i) removed duplicate sequences and computed raw tRI signals with EVOtRec, and (ii) reconstructed phylogenetic trees with IQ-TREE v3.0.1 \cite{Wong2025} and subsequently computed phylogenetic homoplasy counts with TreeTime~v0.11.4~\cite{Sagulenko2018}.
The scatter plot shows mean raw tRI and mean TreeTime homoplasy counts per SNV (means per SNV calculated by averaging over all 100 simulations).
We found that (i) there were two clusters corresponding to whether the SNV was neutral or beneficial, (ii) mean raw tRI signals were smaller than mean TreeTime homoplasy counts, and (iii) several neutral SNVs exhibited no mean raw tRI signal but had positive TreeTime homoplasy count.
Moreover, raw tRI profiles and TreeTime homoplasy counts showed a moderate-to-strong mean Spearman correlation of $\rho = 0.59 \pm 0.01$ across beneficial variants (mean ± 95\% CI, calculated by averaging over all 100 simulations), but they were not correlated across neutral variants ($\rho = 0.04 \pm 0.007$).
}
\label{fig:simulated_tri_multiplicity}
\end{figure}

\begin{figure}[h!]
\centering
\includegraphics[width=\textwidth]{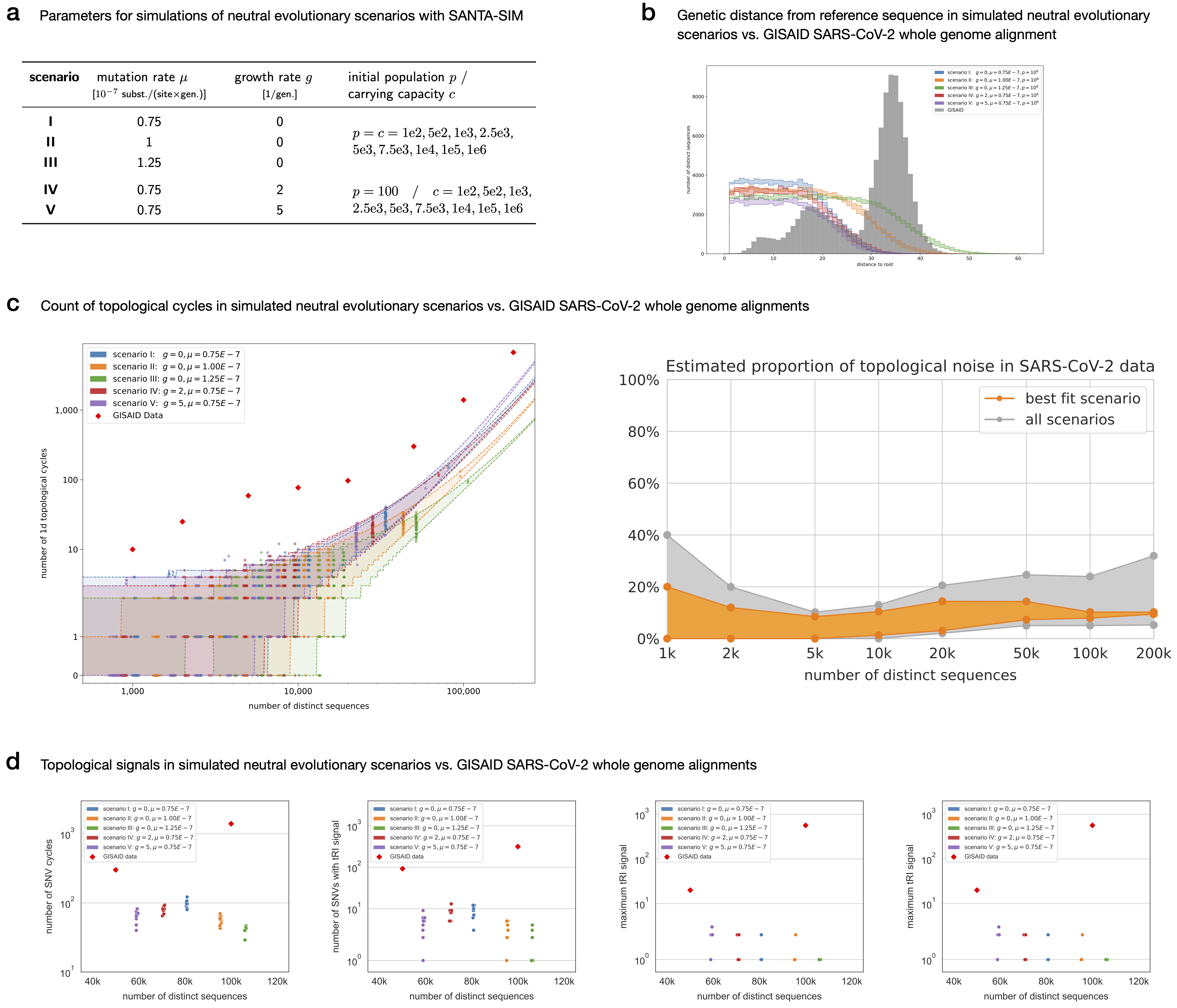}
\caption{\textbf{\sffamily{Robustness of tRI signals to topological noise.}}
We estimated topological background noise in SARS-CoV-2 whole genome alignments from topological signals in neutral evolutionary scenarios, based on synthetic alignments generated with SANTA-SIM v1.0 \cite{Jariani2019} for five distinct scenarios with varying growth rate $g$ and mutation rate $\mu$ (\nameref{methods}).
\textbf{(a)} SANTA-SIM parameter choices for the five simulated scenarios.
Scenarios I-III vary over a range of mutation rates, while scenarios~IV and~V probe the effect of logistic population growth.
\textbf{(b)} Comparison of genetic distances to the root in simulated alignments vs.~genetic distances to the reference sequence hCoV-19/Wuhan/WIV04/2019 in the GISAID \cite{Elbe2017} SARS-CoV-2 200k whole genomes alignment. 
The alignment of scenario~II best fits the SARS-CoV-2 data.
\textbf{(c)} We computed topological cycles with Ripser \cite{Bauer2021} in GISAID SARS-CoV-2 whole genome alignments of size ranging from 1k up to 200k sequences, and in simulated alignments of varying size for the five distinct scenarios.
The 95\% prediction intervals for the number of cycles in each scenario are based on the extrapolation of a Panjer distribution for an increasing number of distinct sequences in simulated phylogenies.
For each scenario, the validation dataset shown in the plot is well-described by the corresponding prediction intervals.
We used the results in (c) to estimate the proportion of topological noise (95\% confidence bands) in SARS-CoV-2 whole genome alignments.
The best fit scenario is scenario~II.
\textbf{(d)} Summary statistics of topological signals in simulated alignments for scenarios I-IV and in GISAID SARS-CoV-2 whole genome alignments with 50k and 100k sequences.
Scatter plots showing the number of SNV cycles, the number of SNVs with (positive) tRI signal, the total sum of tRI signals and the maximum tRI signal in simulated vs.~SARS-CoV-2 alignments (\nameref{methods}).
The number of distinct sequences may vary across alignments because we removed duplicate sequences from the original simulated alignments, each comprising 200k sequences.
}
\label{fig:topological_noise}
\end{figure}

\begin{figure}[h!]
\centering
\includegraphics[width=\textwidth]{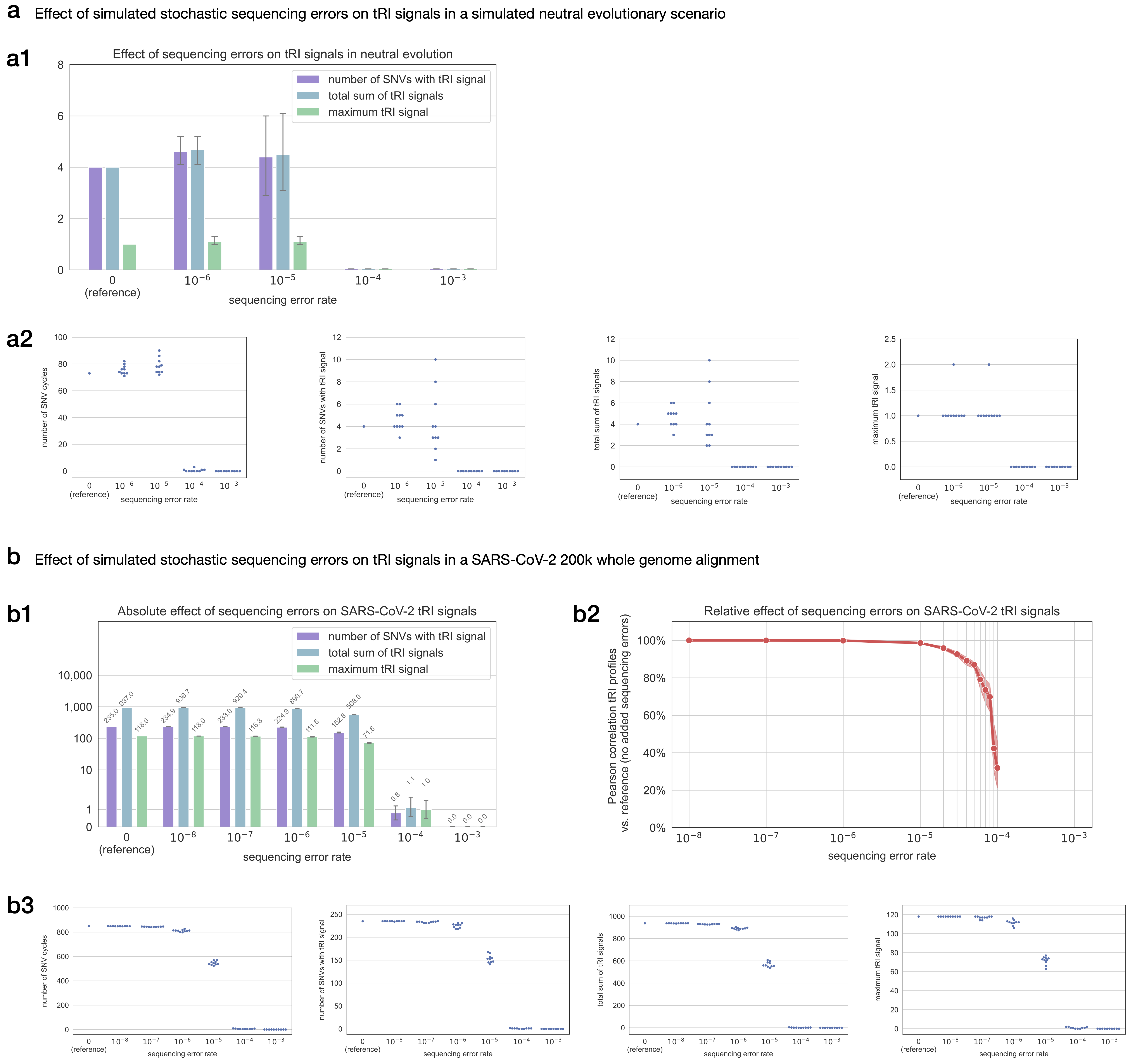}
\caption{\textbf{\sffamily{Robustness of tRI signals to stochastic sequencing errors.}}
We estimated the effect of stochastic sequencing errors on tRI by adding simulated stochastic sequencing errors to two reference alignments: a simulated neutral evolutionary scenario~(a) generated with SANTA-SIM v1.0 \cite{Jariani2019}, and a real SARS-CoV-2 200k whole genomes alignment~(b) (\nameref{methods}).
We analyzed a wide range of error rates from $10^{-8}$ to $10^{-3}$ on a logarithmic scale.
Rates of stochastic sequencing errors in consensus genomes typically lie within this range \cite{Bull2020, Stoler2021}.
We generated 10 alignments with added simulated sequencing errors for each scenario.
\textbf{(a1)} Bar plot showing the absolute effect of sequencing errors on tRI (mean ± 95\% CI) in neutral evolution.
\textbf{(a2)} Swarm plots showing the number of SNV cycles and topological signals (as in a1) in the reference alignment and in alignments with added sequencing errors.
\textbf{(b1)} Bar plot showing the absolute effect of added sequencing errors on tRI (mean ± 95\% CI; bar value labels indicate means) for the original alignment without added sequencing errors (reference) and alignments with added sequencing errors.
The presence of sequencing errors does not increase tRI signals, but decreases tRI signals for larger error rates.
\textbf{(b2)} Line plot showing the Pearson correlation (mean ± 95\% confidence band) between tRI profiles across the whole genome for alignments with added sequencing errors vs.~those for the reference alignment.
\textbf{(b3)} Swarm plots showing the number of SNV cycles and topological signals (as in b1 and b2) in the reference alignment and in alignments with added sequencing errors.
}
\label{fig:sequencing_errors}
\end{figure}

\begin{figure}[h]
  \centering
	\includegraphics[width=\textwidth]{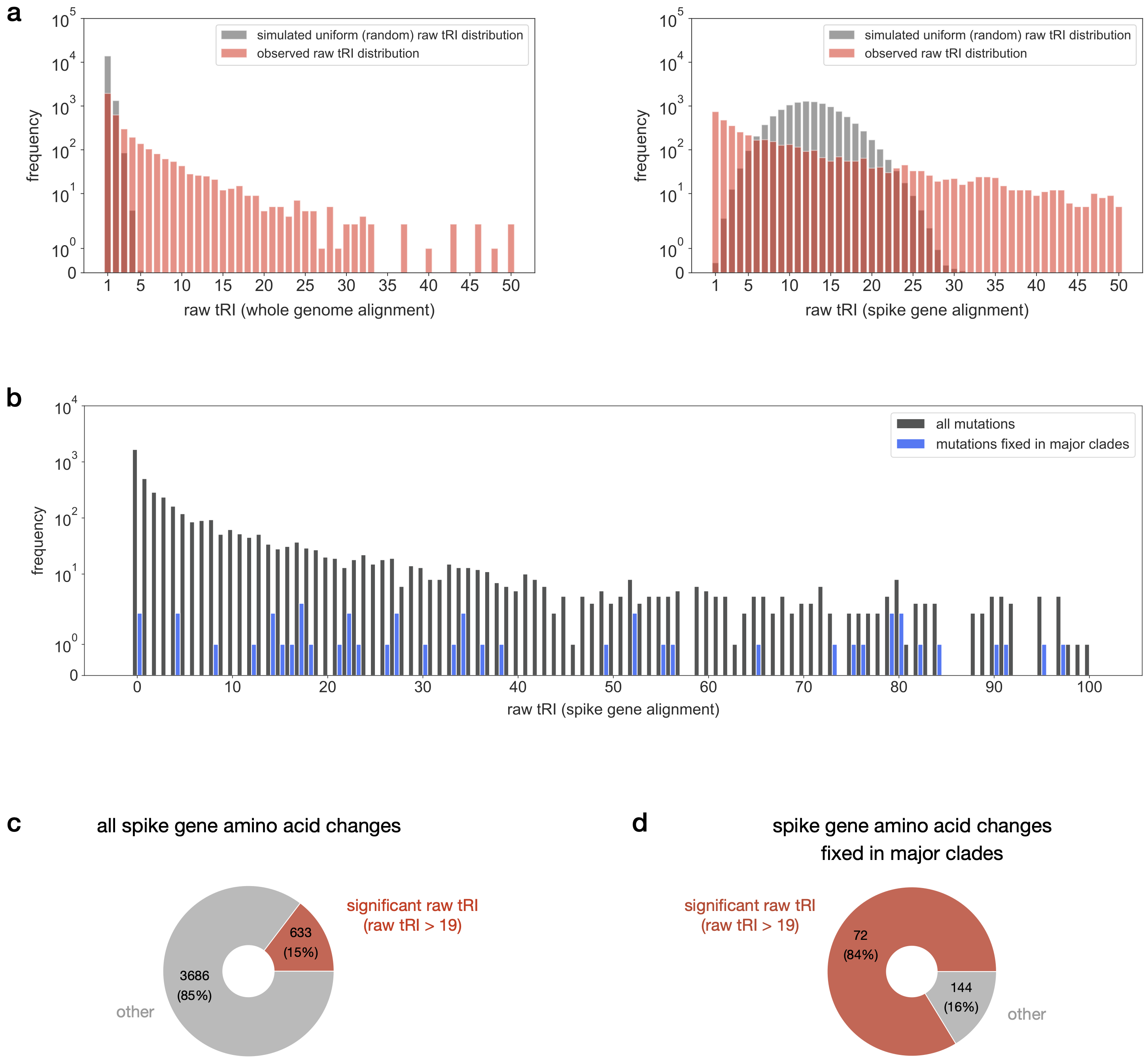}
	\caption{\textbf{\sffamily{Statistical significance of SARS-CoV-2 tRI signals.}}
Raw tRI signals of the SARS-CoV-2 whole genome and spike gene alignments.
\textbf{(a)} The frequency distributions of raw tRI signals in the whole genome alignment and spike gene alignment differ significantly from the frequency distributions obtained from a Monte Carlo permutation test under the null hypothesis that raw tRI signals are distributed uniformly across the genome (Kolmogorov–Smirnov test, $p<0.05$; \nameref{methods}).
In the diagrams only raw tRI signals $\le 50$ are shown.
\textbf{(b)} Frequency distributions of raw tRI signals for all 4319 amino acid changes in the spike gene alignment, and for the subset of those 86 amino acid changes that became fixed in one of the Nextstrain clades Alpha (20I), Beta (20H), Gamma (20J), Delta (21A, 21I, 21J) or Omicron (21K, 21L, 22A, 22B, 22C, 22D, 22E, 22F, 23A, 23B, 23C, 23D, 23E, 23F, 23G, 23H) \cite{Hadfield2018, Hodcroft2024}.
Only raw tRI signals $\le 100$ are shown.
\textbf{(c-d)} Proportion of spike amino acid changes with significant raw tRI signal ($\textrm{raw tRI} > 19$) in all amino acid changes (c) and in amino acid changes fixed in one of the Nextstrain clades Alpha, Beta, Gamma, Delta or Omicron (d).
}
\label{fig:tri_frequency_distribution}
\end{figure}

\begin{figure}[h]
  \centering
	\includegraphics[width=\textwidth]{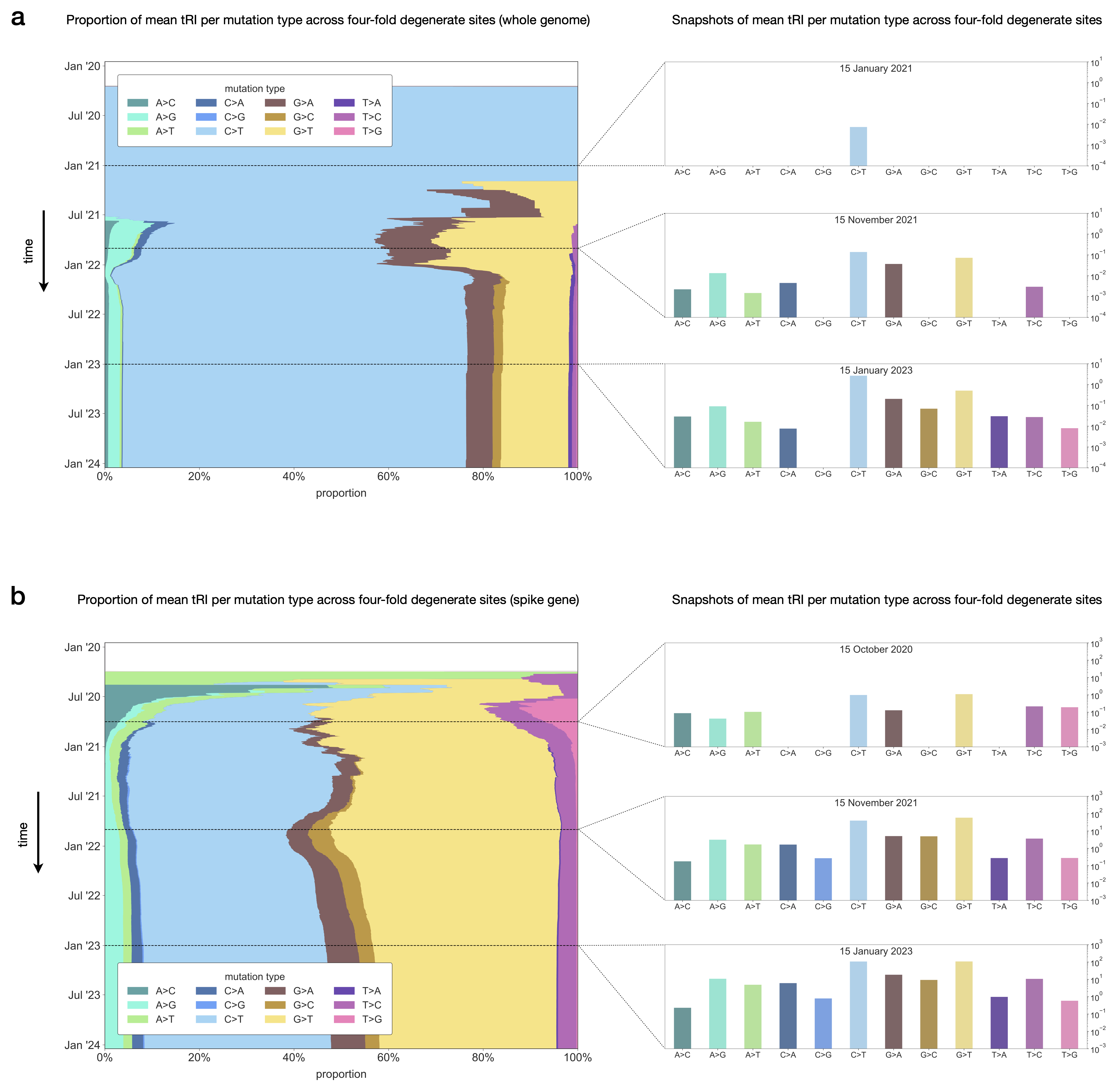}
	\caption{\textbf{\sffamily{Proportions of tRI signals per nucleotide mutation type across four-fold degenerate sites in SARS-CoV-2.}}
Stacked area plots (left) showing the time-dependent relative proportion of mean tRI signals across all four-fold degenerate sites for each nucleotide mutation type on the whole genome (a) and on the Spike gene (b). 
Snapshot bar plots (right) taken at specific time points in the early, middle and late phase of the pandemic showing the absolute mean tRI across all four-fold degenerate sites for each nucleotide mutation type.
Proportions from mid-2022 onwards in (a) are congruent with proportions obtained from mutation counts in SARS-CoV-2 phylogenetic trees \cite{Bloom2023a}.
}
\label{fig:tri_fds}
\end{figure}

\begin{figure}[h]
  \centering
 	\includegraphics[width=\textwidth]{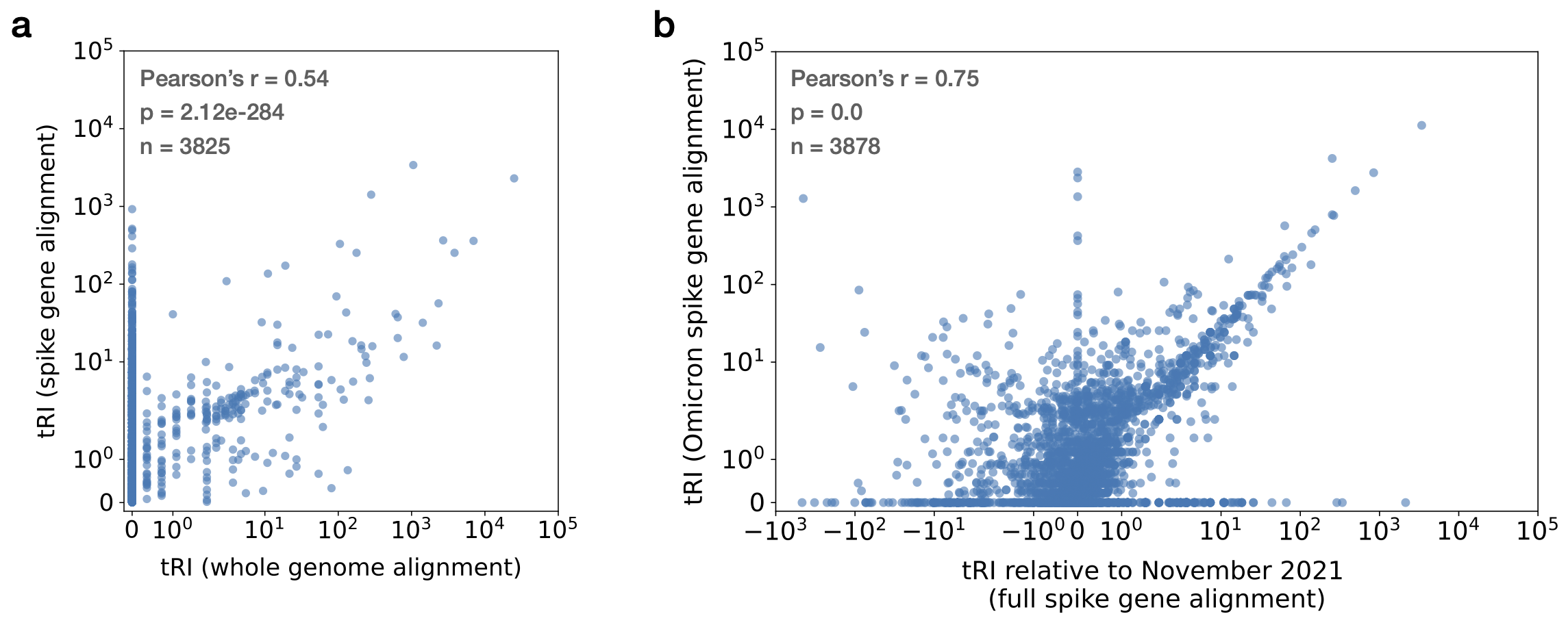}
	\caption{\textbf{\sffamily{Restricting the topological recurrence analysis to sliding windows or sub-alignments.}}
\textbf{(a)} The topological recurrence analysis can be carried out for sliding windows on the genome, such as single genes or specific genomic regions (\nameref{methods}).
A comparison of tRI signals for SARS-CoV-2 spike gene SNVs obtained from the analysis of the whole genome alignment vs.~the spike gene alignment illustrates the trade-off between tRI signal strength and genomic scope.
Restricting the analysis to the spike gene alignment means that SNVs outside the spike gene do not contribute to genetic distance, which typically leads to the creation of more SNV cycles and hence stronger tRI signals.
We found that tRI signals of the whole genome alignment and spike gene alignment are correlated with Pearsons's $r = 0.54$.
\textbf{(b)} Since isolates in SNV cycles share the same genomic background, contributions to tRI signals can be attributed to unique genomic lineages (\autoref{fig:DefinitionOftRI}).
For example, we were able to extract tRI signals generated by SARS-CoV-2 Omicron sequences directly from tRI signals for the full spike gene alignment without the need to run the tRI analysis separately for a sub-alignment corresponding to the Omicron variant.
This was achieved by counting only those tRI signals in the full spike gene alignment that arose after the onset of Omicron in November~2021.
We found that tRI signals occurring after October~2021 in the full spike gene alignment (covering the period from January~2020 until February~2024) and tRI signals in the Omicron spike gene alignment (covering the period from November~2021 until February~2024) are strongly correlated with Pearson's $r = 0.75$.
}
\label{fig:tri_region_subalignment}
\end{figure}

\begin{figure}[h]
  \centering
	\includegraphics[width=\textwidth]{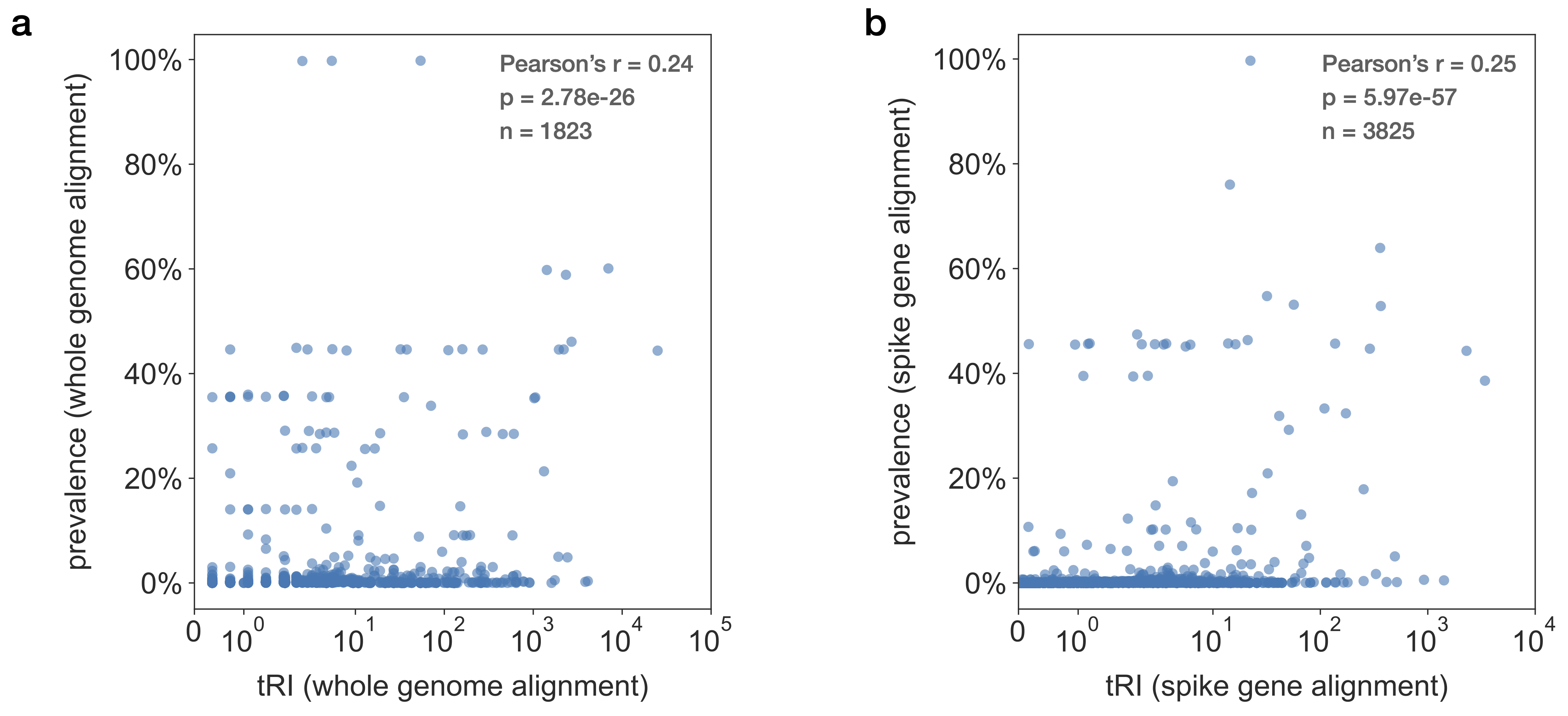}
	\caption{\textbf{\sffamily{Correlation between tRI and prevalence.}}
Scatter plots showing tRI vs.~prevalence of SNVs on the last day of the whole genome alignment (a) and spike gene alignment (b) tRI/prevalence time series (29/30 January 2024).
We found that tRI and prevalence were correlated with Pearson's $r = 0.24$ for the whole genome alignment and $r = 0.25$ for the spike gene alignment.
}
\label{fig:tri_prevalence}
\end{figure}

\begin{figure}[h]
  \centering
	\includegraphics[width=\textwidth]{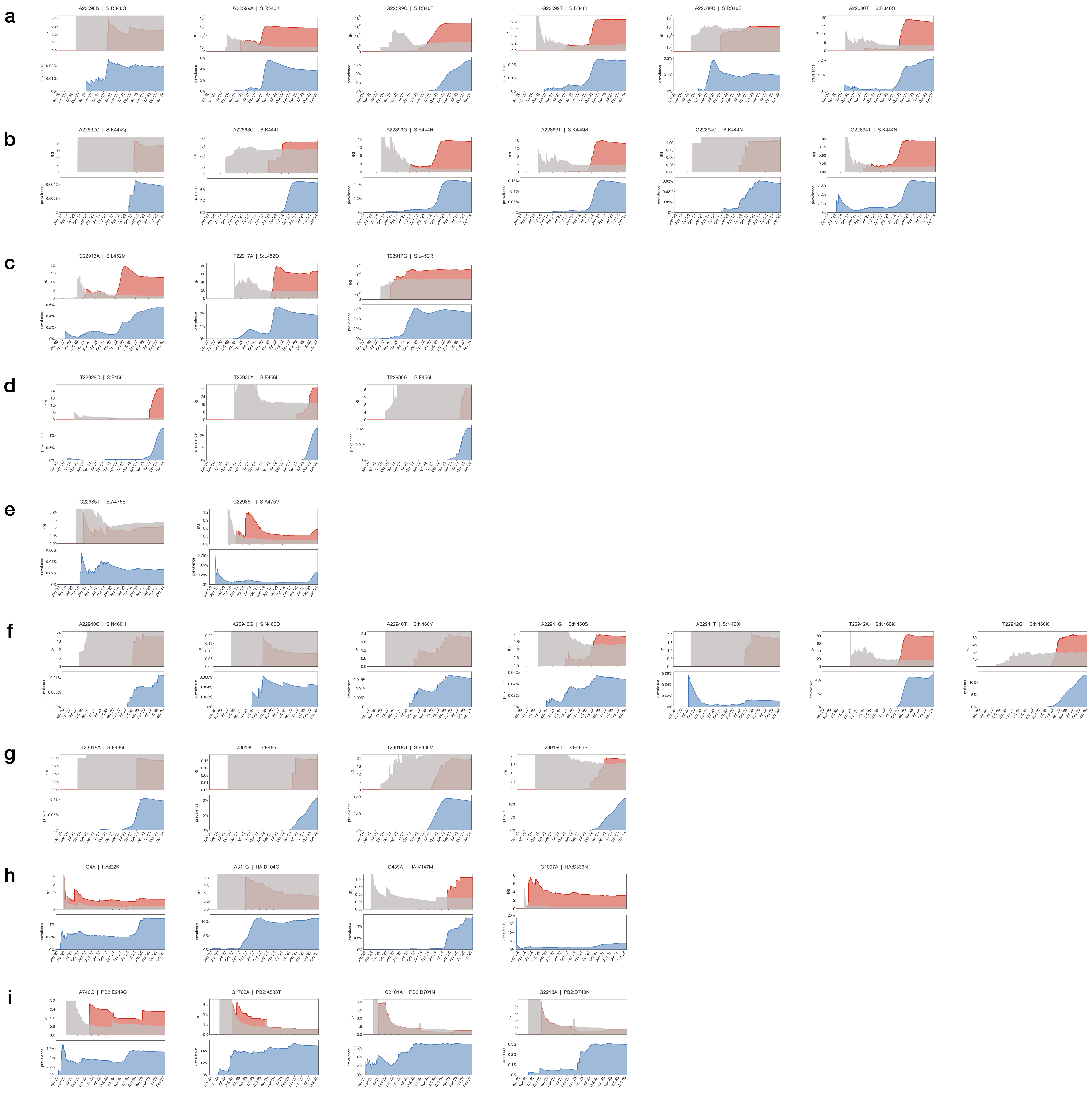}
	\caption{\textbf{\sffamily{Time series tRI signals of selected SARS-CoV-2 and H5N1 amino acid changes.}}
\textbf{(a-g)} Panels displaying tRI/prevalence time series charts of selected SNVs giving rise to SARS-CoV-2 spike gene amino acid changes in Omicron sub-lineages at sites R346 (a), K444 (b), L452 (c), F456 (d), A475 (e), N460 (f) and F486 (g), showing tRI (red; area shaded in grey marks tRI significance level) and prevalence (blue) at daily resolution.
Charts were generated with EVOtRec from the spike gene alignment covering the entire COVID-19 pandemic.
We observed significant tRI for at least one amino acid change at each of these sites, indicating convergent evolution.
This is in line with published studies \cite{Cao2022, Ito2023, Jian2024} which demonstrated that amino acid changes at the sites reported in (a-g) are convergent and were acquired recurrently across several Omicron sub-lineages.
\textbf{(h-i)} Panels displaying tRI/prevalence time series charts of selected SNVs giving rise to H5N1 HA and PB2 gene amino acid changes, showing tRI (red; area shaded in grey marks tRI significance level) and prevalence (blue) at daily resolution.
(h) The HA mutations E2K, D104G, V147M and S336N have been associated with antibody neutralization in bovine H5N1 \cite{Miyakawa2025}; we found significant tRI signals for E2K, V147M and S336N.
(i) Significant tRI signals of the mammalian-adaptive PB2 mutations E249G \cite{Yamaji2015}, A588T \cite{Fan2014}, D701N \cite{Gao2009, Peacock2023} and  D740N \cite{Dholakia2026}.
}
\label{fig:tri_convergent_mutations}
\end{figure}

\begin{figure}[h]
  \centering
	\includegraphics[width=\textwidth]{figures/figure_stats_tri_fitness_dms.png}
	\caption{\textbf{\sffamily{SARS-CoV-2 topological signals correlate with positive selection in experimental data from deep mutational scanning.}}
\textbf{(a-d)} Correlation between EVOtRec tRI signals of SARS-CoV-2 spike amino acid changes in the Omicron spike gene alignment (\nameref{methods}) and published pseudovirus deep mutational scanning (DMS) measurements \cite{Dadonaite2023, DadonaiteDMS2023, Dadonaite2024a, DadonaiteDMS2024ba2, DadonaiteDMS2024xbb} for various phenotypes of the BA.1 (a), BA.2 (b) and XBB.1.5 (c) sub-lineages.
In each case in (a-c), we tested the performance of the tRI in identifying fitness-increasing amino acid changes (i.e.~positive DMS effect).
For this, we analyzed correlations statistically using contingency tables for binarized tRI ($\textrm{tRI} > 0$ vs.~$\textrm{tRI} = 0$) and binarized DMS effect ($\textrm{DMS effect} > 0$ vs.~$\textrm{DMS effect} \le 0$) (a-c) and on a logistic regression analysis for the three phenotypes of XBB.1.5 (d) (\nameref{methods}).
For comparison with phylogeny-based methods, all panels include similar statistical analyses for binarized fitness effect estimates ($\textrm{fitness effect} > 0$ vs.~$\textrm{fitness effect} \le 0$) from \cite{Bloom2023a, Haddox2025} using the datasets \cite{BloomNeherData2024b, HaddoxData2025_21K} for BA.1 (a), \cite{BloomNeherData2024b, HaddoxData2025_BA2} for BA.2 (b) and \cite{BloomNeherData2024b, HaddoxData2025_XBB} for an ordinary least squares regression analysis for XBB.1.5 (c) (\nameref{methods}).
\textbf{(e)}~Comparison of the performance of tRI vs.~phylogeny-based fitness effect estimates \cite{Bloom2023a, Haddox2025} in identifying fitness-increasing amino acid changes (i.e.~positive DMS effect), based on the results from (a-c).
}
\label{fig:stats_tri_fitness_dms}
\end{figure}

\begin{figure}[h]
  \centering
	\includegraphics[width=10cm]{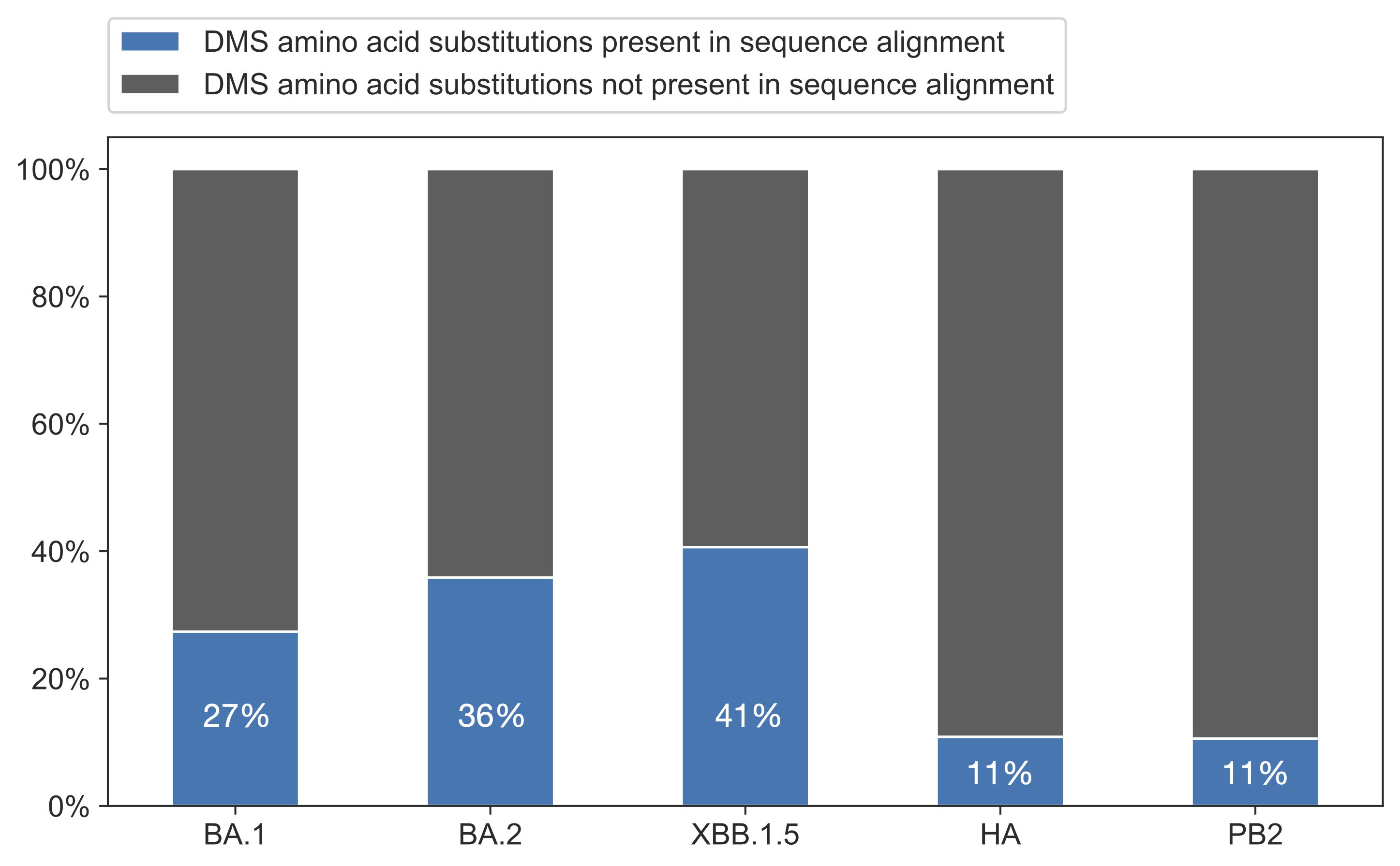}
	\caption{
Percentages of amino acid changes that have been tested in the DMS experiments in \cite{Dadonaite2023, Dadonaite2024a, Dadonaite2024b, Soh2019} for the SARS-CoV-2 Omicron spike gene alignment (covering the sublineages BA.1, BA.2 and XBB.1.5) and the two avian influenza H5N1 alignments (covering the HA and PB2 genes).
}
\label{fig:stats_mutations_in_dms}
\end{figure}

\begin{figure}[h]
  \centering
	\includegraphics[width=\textwidth]{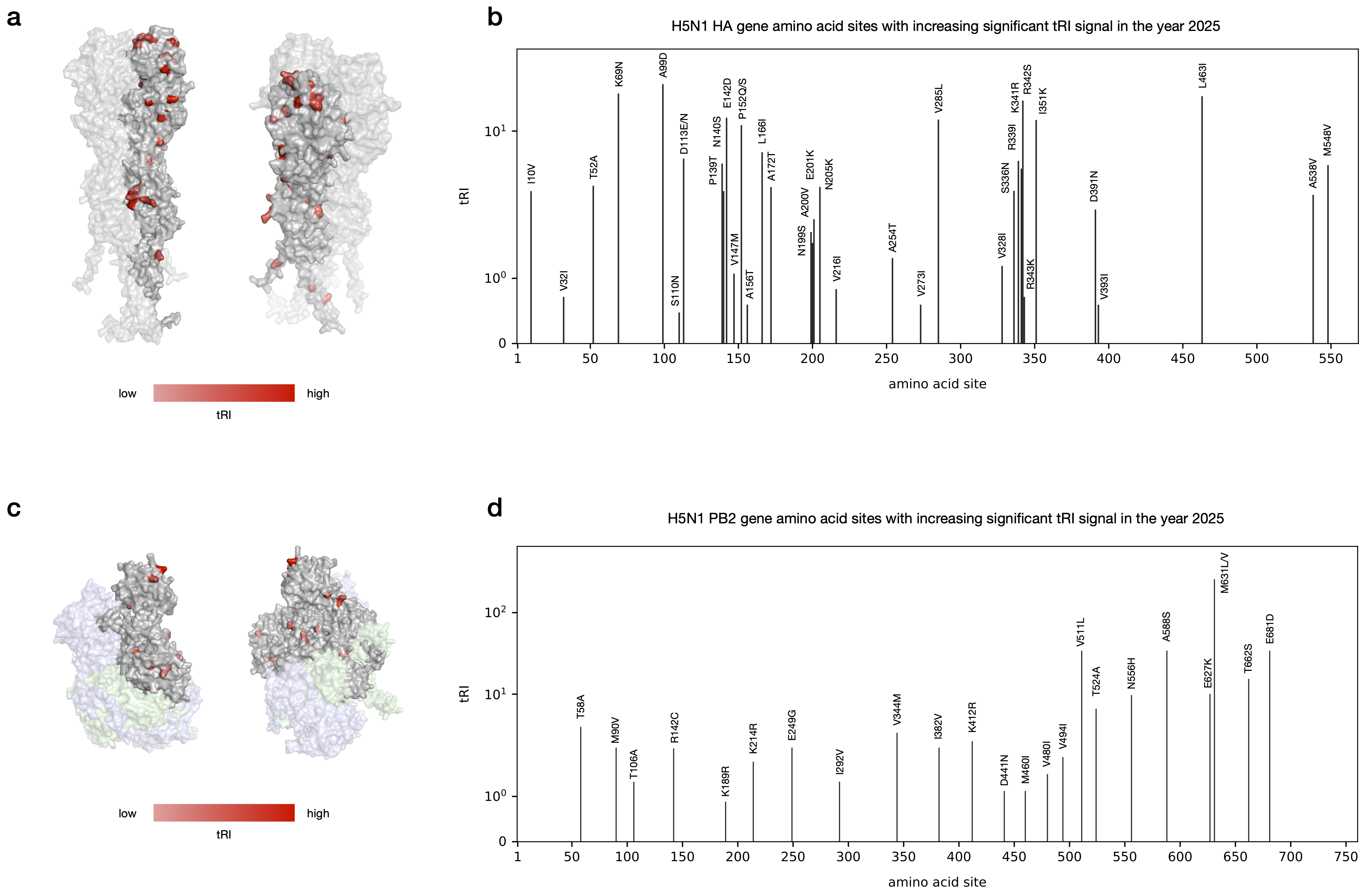}
	\caption{\textbf{\sffamily{Convergent H5N1 amino acid changes.}}
Amino acid changes on the H5N1 HA and PB2 genes that exhibited increasing significant tRI signal during the year 2025. Displayed tRI scores are taken from the last day of the HA and PB2 tRI time series (7/8 November 2025) (\hyperref[tab:supp_tri_influenza_ha]{Supplementary Table~\ref{tab:supp_tri_influenza_ha}} and \nameref{methods}).
\textbf{(a, c)} Two views of the H5N1 HA (a) and PB2 (c) proteins, with residues colored by sum of tRI scores across observed changes at the site. The HA protein is a homo-trimer; one sub-unit is shown with higher opacity and has the recurrently mutated residues highlighted. The polymerase complex is a hetero-trimer consisting of PB1 (pale green), PA (pale blue), and PB2 (light grey). Protein structures were predicted with AlphaFold3 \cite{Abramson2024} as homo-trimer (HA) and heterotrimer (PB2) based on the reference sequence A/American\_Wigeon/South\_Carolina/22-000345-001/2021, and visualized using PyMOL \cite{DeLano2002}.
\textbf{(b, d)} Bar plots of amino acid changes and their tRI signals across the HA and PB2 gene.
}
\label{fig:h5n1_convergent_mutations}
\end{figure}

\end{document}